\documentclass[fleqn,usenatbib]{mnras}

\usepackage{newtxtext,newtxmath}

\usepackage[T1]{fontenc}

\DeclareRobustCommand{\VAN}[3]{#2}
\let\VANthebibliography\thebibliography
\def\thebibliography{\DeclareRobustCommand{\VAN}[3]{##3}\VANthebibliography}

\usepackage{graphicx}	
\usepackage{amsmath}	
\usepackage{mathtools} 
\usepackage{longtable}
\usepackage{caption}
\usepackage{orcidlink}

\defcitealias{Poggianti2016}{P16}
\defcitealias{Vulcani2022}{V22}
\defcitealias{Smith2022}{S22}

\title[Optical jellyfish tail directions in clusters
]{Constraining the duration of ram pressure stripping features in the optical from the direction of jellyfish galaxy tails
}

\author[Salinas et al.]{Vicente Salinas$^{1,2} \orcidlink{0000-0002-6858-4976}$ \thanks{E-mail: froemel@chalmers.se}, Yara L. Jaff\'e$^{3}$ \orcidlink{0000-0003-2150-1130} \thanks{E-mail: yara.jaffe@usm.cl},
Rory Smith$^{3} \orcidlink{0000-0001-5303-6830}$, 
Jong-Ho Shinn$^{4}$ \orcidlink{0000-0001-7967-6473},
Jacob P. Crossett$^{2,3}$ \orcidlink{0000-0002-9810-1664}, 
\newauthor
Marco Gullieuszik$^{5}$,
Gemma Gonz\'alez-Tor\`a$^{9,10}$,
Franco Piraino-Cerda$^{3}$ \orcidlink{0009-0008-0197-3337}, 
Bianca Poggianti$^{5}$, 
\newauthor
Benedetta Vulcani$^{5}$ \orcidlink{0000-0003-0980-1499}, Andrea Biviano$^{6,7}$\orcidlink{0000-0002-0857-0732},
Ana C. C. Louren\c{c}o$^{2,8}$ \orcidlink{0000-0002-4393-7798}, 
Lawrence E. Bilton$^{2,3,11,12}$ \orcidlink{0000-0002-4780-129X},
\newauthor
Kshitija Kelkar$^{2,3}$,
Paula Calder\'on-Castillo$^{3}$
\\
$^{1}$Department of Space, Earth and Environment, Chalmers University of Technology, Onsala Space Observatory, 43992 Onsala, Sweden\\
$^{2}$Instituto de F\'isica y Astronom\'ia, Universidad de Valpara\'iso, Avda. Gran Breta\~na 1111, Valpara\'iso, Chile\\
$^{3}$Departamento de F\'isica, Universidad T\'ecnica Federico Santa Mar\'ia, Avenida Espa\~na 1680, Valpara\'iso, Chile\\
$^{4}$Korea Astronomy and Space Science Institute (KASI), 776 Daedeok-daero, Yuseong-gu, Daejeon 34055, Republic of Korea\\
$^{5}$INAF-Osservatorio Astronomico di Padova, vicolo dell’Osservatorio 5, 35122 Padova, Italy\\
$^{6}$INAF-Osservatorio Astronomico di Trieste, via G. Tiepolo 11, 34143 Trieste, Italy\\
$^{7}$IFPU-Institute for Fundamental Physics of the Universe, via Beirut 2, 34014 Trieste, Italy \\
$^{8}$European Southern Obervatory (ESO), Alonso de Cordova 3107, Santiago, Chile \\
$^{9}$Zentrum für Astronomie der Universität Heidelberg, Astronomisches Rechen-Institut, Mönchhofstr. 12-14, 69120 Heidelberg \\
$^{10}$European Southern Observatory (ESO), Karl Schwarzschildstrasse 2, 85748 Garching bei München, Germany \\
$^{11}$Instituto de Física, Pontificia Universidad Católica de Valparaíso, Casilla 4059, Valparaíso, Chile
 \\
$^{12}$Centre of Excellence for Data Science, Artificial Intelligence \& Modelling, The University of Hull, Cottingham Road, Kingston-Upon-Hull, HU6 7RX, UK
}

\date{Accepted 2024 July 18. Received 2024 July 3; in original form 2023 September 22}

\pubyear{2024}

\begin{document}
\label{firstpage}
\pagerange{\pageref{firstpage}--\pageref{lastpage}}
\maketitle

\begin{abstract}
Ram pressure stripping is perhaps the most efficient mechanism for removing gas and quenching galaxies in dense environments as they move through the intergalactic medium. Extreme examples of on-going ram pressure stripping are known as jellyfish galaxies, characterized by a tail of stripped material that can be directly observed in multiple wavelengths. Using the largest homogeneous broad-band optical jellyfish candidate sample in local clusters known to date, we measure the angle between the direction of the tails visible in the galaxies, and the direction towards the host cluster center. We find that $33\%$ of the galaxy tails point away from the cluster center, $18\%$ point towards the cluster center, and $49\%$ point elsewhere. Moreover, we find stronger signatures of ram pressure stripping happening on galaxies with a tail pointing away and towards the cluster center, and larger velocity dispersion profiles for galaxies with tails pointing away. These results are consistent with a scenario where ram pressure stripping has a stronger effect for galaxies following radial orbits on first infall. The results also suggest that in many cases, radially infalling galaxies are able to retain their tails after pericenter and continue to experience significant on-going ram pressure stripping. We further constrain the lifespan of the optical tails from the moment they first appear to the moment they disappear, by comparing the observed tail directions with matched N-body simulations through Bayesian parameter estimation. We obtain that galaxy tails appear for the first time at $\sim 1.16$ R$_{200}$ and disappear $\sim660$ Myr after pericenter.

\end{abstract}

\begin{keywords}
galaxies: clusters: general -- galaxies: clusters: intracluster medium -- galaxies: evolution
\end{keywords}



\section{Introduction}
\label{sec:Introduction} 

Its been established by numerous studies that galaxies evolve differently depending on their mass and their environment \citep[e.g.][]{Davies&Lewis1973,
Dressler1980,
Kennicutt1983,
Giovanelli&Haynes1985,
Whitmore1993, 
Solanes2001,
Baldry2006,
Gavazzi2010,
Peng2010}. 
These (and other) studies find that the most massive galaxies tend to be evolved passive systems regardless of their environment, but lower mass systems have properties which depend strongly on environment. In essence the vast majority of galaxies can loose their ability to form stars,  "age", and transform morphologically as they migrate from low-density regions of the cosmic web to increasingly higher density environments (like groups and clusters of galaxies).

Different physical mechanisms have been proposed to explain the large fractions of passive galaxies in dense environments \citep[see][for a complete review]{Boselli&Gavazzi2006,Cortese2021,Boselli2022}. 
In general, environmental effects can be divided in two main groups: \textbf{i)} gravitational interactions, such as galaxy-galaxy mergers, galaxy-cluster interactions, or harassment \citep{Spitzer&Baade1951,Merritt1983,Moore1998}; and \textbf{ii) }hydrodynamical effects, such as ram pressure stripping (RPS), thermal evaporation, and viscous stripping \citep{Gun&Gott1972,Cowie&Songalia1977,Nulsen1982}, caused by the interaction between the cold interstellar medium (ISM) of galaxies and the hot and dense intracluster medium (ICM) present in the cluster.  The main difference between gravitational and hydrodynamical effects is that the former is capable of perturbing all the components of the galaxy, while the latter can only directly perturb the gas content. However, both effects are capable of stripping/consuming large quantities of gas from galaxies, helping them transform and ultimately quench their star formation. 

RPS is thought to be among the most effective mechanisms altering the gas content of galaxies in clusters \citep{LeeSeona2022}: 
As galaxies fall into clusters, they experience a drag pressure capable of stripping some (or all) of the gas in the ISM, analytically described by \citet{Gun&Gott1972} as: 
\begin{equation}
    \centering
    P_{\text{ram}} \approx \rho_{\text{e}} v^{2} \text{,}
    \label{eq:G&G}
\end{equation}
where  $\rho_{\text{e}}$ is the density of the ICM, and $v$ is the velocity of the galaxy relative to the medium. If $P_{\rm ram}$ overcomes the binding force of the galaxy, gas will be removed from the body of the galaxy leaving the underlying stellar component unaltered (albeit not completely, see \citeauthor{Smith2012} \citeyear{Smith2012}).

Several studies have shown that RPS can significantly remove gas from galaxies on their first infall into the clusters and that radial orbits provide more intense stripping \citep{Dressler1986,Giraud1986,Solanes2001,Jaffe2015,Jaffe2018,Yoon2017}. In addition, radio observations show clear signs of disturbed and stripped gas with the main stellar body of the galaxy unaffected, as expected with RPS \citep{Warmels1988_3,Cayatte1990,Gavazzi1995,Chung2009} sometimes accompanied by recent star formation in the stripped tails \citep[e.g][]{Gavazzi1995,Gullieuszik2023} for which these galaxies became known as "jellyfish" galaxies.

Although individual examples of  jellyfish galaxies  have been observed at different wavelengths \citep{Sun2006,SmithRu2010,Ebeling2014,Kenney2014,Fossati2016}, large surveys of "jellyfish" or RPS candidates began to appear recently, starting with the works of \cite{McPartland2016} and \citet[][hereafter \citetalias{Poggianti2016}]{Poggianti2016} who provided large catalogues of RPS candidates selected from broad-band optical images. With these works, we began to more confidently define some of the common properties of these galaxies, and in turn, better understand the effects of RPS. In particular, these samples show that $15-25$\% of infalling cluster galaxies show visible signs of stripping on optical broad band images, which are only the tip of the iceberg \citep[][see also Crossett et al. submitted]{Merluzzi2013,Vulcani2022,Pedrini2022,Lourenco2023}. 

The large sample of RPS candidates from \citetalias{Poggianti2016} also gave rise a follow-up with integral-field spectroscopy  \citep[the GASP MUSE survey][]{Poggianti2017}, providing unprecedented detail into the formation of these galaxies and the impact of RPS on galaxy evolution. 
GASP found, among many other things, evidence of outside-in gas stripping on infalling cluster galaxies, which generates tails of debris material opposite to the direction of motion. These tails often display knots of newly formed stars (outside of the galaxy) and, in some cases, signs of unwinding of spiral arms \citep{Bellhouse2021}. At the same time, some cases of intensely stripped galaxies also show nuclear activity suggesting the interaction with the ICM can help feed supermassive black holes \citep[][]{Poggianti2017agn, Gullieuszik2017,Vulcani2018b,Poggianti2019,Bellhouse2017,Bellhouse2021}.

One of the pending questions regarding the effectiveness of RPS in the quenching of galaxies \citep{Wetzel2013,Haines2015} is the timescale of stripping \citep[e.g.][]{Schultz&Struck2001,Roediger&Hensler2005,Steinhauser2016,Smith2022,Rohr2023}. To constrain the time it takes for galaxies to loose their gas as they fall and virialize into clusters, its essential to understand the orbits of the galaxies. These can be reconstructed using position-velocity phase-space information \citep[e.g.][]{Jaffe2015,Jaffe2016,Pasquali2019}, and/or the direction of the stripped tails, in conjunction with cosmological simulations as explained below. 

Given that the tails of RPS galaxies are expected to point opposite to the direction of motion, we can use them to obtain direct information about their orbits. One of the first studies to mention the tail orientation of multiple jellyfish galaxies in a cluster is the work by \cite{Chung2007}, where they find that most galaxies with HI tails in the Virgo cluster point away from the cluster center. This behaviour has been confirmed by tailed cluster galaxies seen in UV \citep{SmithRu2010}, radio-continuum \cite{Roberts&Parker2020}, but naturally not seen as clearly in merging clusters where the orbits are more complex \cite{Rawle2014,Roman-Oliveira2021}. 
The prevalence of tails pointing away from the cluster centers (in virialized systems) is consistent with a scenario where tailed galaxies are seen as they fall into the cluster for the first time on preferentially radial orbits, an interpretation that is supported by the phase-space distribution of jellyfish galaxies \citep[][]{Jaffe2018}, as well as simulations \citep{Smith2022}.

Tail direction studies of optical jellyfish galaxies in multiple clusters began with the large samples that appeared in 2016. On the one hand, \citetalias{Poggianti2016} provided crude estimations of the fractions of the tail orientations, finding $\sim 13\%$ of tails pointing towards, $\sim35\%$ pointing away, and $\sim52\%$ pointing elsewhere. On the other hand, \cite{McPartland2016} made a more extensive study on the direction of motion of their jellyfish candidates, based on the tail directions (although the tail direction distribution is not provided). They used their results in conjunction with hydrodynamical models from \cite{Roediger2006}, \cite{Kronberger2008}, and \cite{Roediger2014} to constrain the infall histories of the galaxies. They find that their distribution best matches a fast cluster merger scenario, rather than galaxy accretion from filaments or slow cluster mergers. However, their results do not rule out contributions from the other scenarios. Their results also agree with a scenario where jellyfish galaxies might also be observed near the cluster center of low-mass clusters or potentially even in groups. 

The recent works of \cite{Roberts2021} and \cite{Roberts2021groups} provide a complete distribution of radio-continuum tail directions from the LOFAR Two-metre Sky Survey (LoTSS) in both clusters and groups, respectively. In the case of clusters, they once again find a distribution that prefers tails pointing away from the cluster center, while in the case of groups, they find a two-peaked distribution, where one peak corresponds to tails pointing away from the group center (but slightly more perpendicular than in clusters) and a secondary peak corresponding to tails pointing towards. A similar two-peaked distribution is found by \cite{Kolcu2022}, using optical tail directions of galaxies in groups from the Galaxy And Mass Assembly (GAMA, \citeauthor{Driver2011} \citeyear{Driver2011}) survey. 
These authors interpret tails pointing away from the cluster center as galaxies infalling towards the cluster center; and tails pointing towards the cluster center as ‘backsplashing’ galaxies moving away from the cluster center. 
Their findings could indicate that infalling galaxies in clusters suffer significant RPS  on first infall, while galaxies in groups experience delayed RPS due to the lower ICM densities and velocities relative to clusters.
 
Recently \cite[][hereafter \citetalias{Smith2022}]{Smith2022} 
developed a novel method to constrain the lifespan of jellyfish galaxy tails using phase-space location of jellyfish galaxies and the distribution of the tail direction angles. In particular, they use N-body cosmological dark matter only simulations, for which they later ``paint-on'' the galaxy tails using free parameters that are constrained by comparing the model with observations through Bayesian parameter estimation. They tested their method on the LoTSS sample from \citet{Roberts2021,Roberts2021groups}, and find that radio continuum tails appear on average at $\sim0.76$ R$_{200}$, and disappear $\sim 480$ Myr after pericenter.

In this work, we present tail direction measurements of a large sample of RPS candidates in clusters selected in the optical, which probes star-formation on a different timescale to radio-continuum. We study the properties of the tailed galaxies and adapt  the  method of \citetalias{Smith2022} to constrain, for the first time, the lifespan of the jellyfish features in the optical regime.  

Although broad-band optical features only show the tip of the iceberg when it comes to RPS, optical images are vastly accessible, allowing for large samples and better statistics. Moreover, optical tails are linked to the process of extraplanar star formation \citep[see e.g.][]{Gullieuszik2020} which makes them an important element to constrain the RPS process. With this work we aim to provide an in-depth study of the directions of optical tails leading to clear constraints in the lifespan of these features, and  hope our methodology will continue to pave the way for future studies that will take advantage of the continuously growing samples of jellyfish galaxies.

Throughout this paper we assume a $\Lambda$CDM cosmology, with a Hubble constant $H_{0} = 70$ $\text{km}$ $\text{s}^{-1}$ $\text{Mpc}^{-1}$, present matter density of $\Omega_{m} = 0.27$, and dark energy density $\Omega_{\Lambda}=0.73$. 

\section{Data and Sample}
\label{sec:data} 

In this work, we use the largest homogeneous sample of optically-selected RPS galaxy candidates in the low-redshift Universe compiled to date, consisting of $379$ galaxies: $344$ of them come from the \citetalias{Poggianti2016} sample, and  $35$ come from the newly identified candidates from \citet[][hereafter \citetalias{Vulcani2022}]{Vulcani2022}. Both of these samples were visually selected from broad-band optical images of clusters from the WINGS and OmegaWINGS surveys  (\citeauthor{Fasano2006} \citeyear{Fasano2006}, \citeauthor{Varela2009} \citeyear{Varela2009}, \citeauthor{Gullieuszik2015} \citeyear{Gullieuszik2015}). In short, WINGS is a survey of $77$ low redshift ($z\sim0.04-0.07$) galaxy clusters selected on the basis of their X-ray luminosity (\citeauthor{Ebeling1996} \citeyear{Ebeling1996}, \citeyear{Ebeling1998}, \citeyear{Ebeling2000}). The WINGS data consists of B and V band photometry plus spectroscopy for most of the clusters, with a typical field of view of $34' \times 34'$, which translates into an average coverage of $\sim 0.8$ R$_{200}$. OmegaWINGS is an extension of WINGS that quadruples the field of view for $46$ of the clusters, yielding an average coverage of $\sim 1.2$ R$_{200}$ for this sub-sample.

In the original RPS candidate sample from \citetalias{Poggianti2016}, they searched for RPS features such as unilateral debris or tails in the optical images. Up to three classifiers assigned to each candidate a "Jellyfish Class" (JClass), which is a visual indication of the strength of the stripping features, going from extreme cases (JClass  $= 5$) to progressively weaker cases, with the weakest case being JClass $= 1$. This sample was expanded by  \citetalias{Vulcani2022} who re-inspected WINGS/OmegaWINGS images to find missed RPS candidates by \citetalias{Poggianti2016}, and also identify cases of "unwinding" spirals that could (or not) be RPS-induced. It is important to note that while \citetalias{Poggianti2016} inspected all cluster galaxies (i.e. the photometric sample), \citetalias{Vulcani2022} only considered spectroscopically confirmed members. 

We do not include the unwinding candidates from \citetalias{Vulcani2022} in our study as they are not confirmed RPS cases, but we note that within the \citetalias{Poggianti2016} sample there are 11 cases of unwinding (identified as such later). These galaxies are part of the GASP survey and have been confirmed to be experiencing RPS by \citet{Bellhouse2021}, who presented a detailed analysis of these galaxies comparing the MUSE data with RPS simulations. 

We note that while some authors reserve the term "jellyfish"  to refer to extreme cases of RPS, in the remaining of this paper we will call all optically-selected RPS candidates "jellyfish" for simplicity. In other words, we have all degrees of stripping in our optical selection, but with strong enough gas removal to cause changes in the optical light (new stars being born outside the regular disk. Although our sample of jellyfish candidates could potentially be contaminated with some galaxies not affected by RPS, we expect low contamination as  
86\% of the RPS candidates from \citetalias{Poggianti2016} observed by GASP are confirmed RPS cases \citep[see][and Poggiantti et al. in prep]{Vulcani2022, Jaffe2018}. Naturally galaxies with lower JClass are more susceptible to be non-RPS cases, but even low JClass candidates have been confirmed to be clear cases of RPS \citep[e.g. the case of JO147][]{Merluzzi2013}. In the following sections we always consider galaxies of any JClass, but in some cases we restrict the sample to  high JClass candidates (at the expense of a larger sample size), in an attempt to minimize the number of missclassifications.

To further reduce the noise in our results, for the present work, we have cleaned the sample of RPS galaxy candidates from possible tidal interactions. We did this by considering the comments in table 3 from \citetalias{Poggianti2016} and dismissed $66$ galaxies with indications of tidal interaction or merger. We also cleaned the sample of clusters from highly interacting clusters. To this end, we used the classifications of the cluster dynamical states described in \cite{Lourenco2023}, for which we consider interacting clusters those flagged as pre-merger, interacting, and post-merger. This yielded $9$ interacting clusters (containing $46$ candidates in total). Lastly, we removed $76$ candidates that are confirmed non-cluster members. After filtering the data we are left with a clean sample of $227$ jellyfish candidates.

In what follows we use stellar masses ($M_{*}$) and redshifts from \citetalias{Vulcani2022}. The total (SEXTRACTOR AUTO) B and V absolute magnitudes for most of the WINGS and OmegaWINGS galaxies (also including non-jellyfish candidates, which we use for reference) are also taken from \citetalias{Vulcani2022}, measured from \cite{Moretti2014} and \cite{Gullieuszik2015}, corrected for distance modulus, foreground galaxy extinction, and k-corrected using tabulated values from \cite{Poggianti1997}. 

Cluster properties (including velocity dispersion, $\sigma_{\text{cl}}$; sizes, $R_{200}$; and host mass, $M_{200}$) come from \cite{Lourenco2023} and \cite{Biviano2017}. Cluster memberships are also taken from \cite{Lourenco2023} which uses the method described in \cite{Biviano2017} and \cite{Paccagnella2017}, based on projected position-velocity phase-space. In our parent sample of jellyfish candidates, we have $195$ confirmed members and $108$ candidates with unknown redshift. Finally, the corresponding BCG for each cluster is taken from the BCG sample in \cite{Fasano2010}, which we treat as the cluster center.

\section{Measuring the tail directions of jellyfish galaxies}

We use the optical images of our jellyfish candidate sample to measure the direction of the tails relative to the center of the clusters. To robustly determine this tail angle for each galaxy, up to seven classifiers (coauthors) inspected the galaxies and interactively drew the tail directions. To maximize the visibility of the (often) low-surface brightness tails, and facilitate the measurement, six different logarithmic min-max scales of the B-band WINGS/OmegaWINGS images were displayed, as well as RGB images from the Legacy Survey \citep[][when available]{Legacy19}. Classifiers were allowed to select and zoom in/out (if necessary) in any of the images to "draw" the tail as a straight line. In addition, each classifier assigned a confidence level based on the clarity of the tails, with possible values of 0 (no tail), 1 (marginal tail), or 2 (clear tail). After the tail is drawn, the direction is computed as the angle with respect to the x-axis of the images in a counterclockwise manner (with the north pointing up and west to the right). 

\begin{figure}
    \centering
    \includegraphics[width=0.48\textwidth]{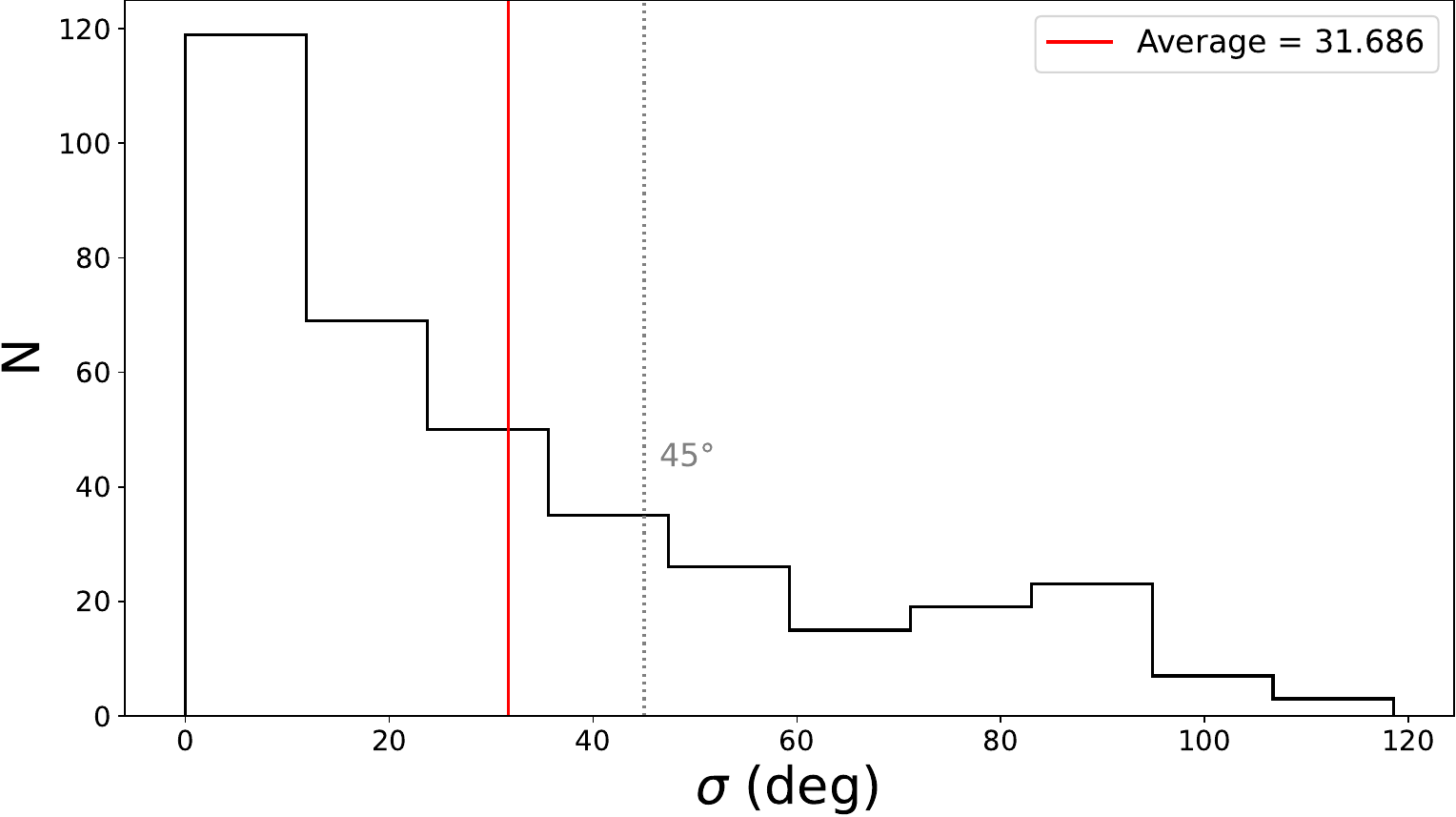}
    \caption{Standard deviation distribution of tail angles for galaxies in which more than one classifier agrees to see a tail. Note that this result serves as an initial test on the agreement of the classifiers, therefore we did not apply the flowchart procedure described in Figure \ref{fig:flowchart}. The method used to compute the standard deviations is described in Appendix~\ref{ap:comparisson}.}
    \label{fig:std_hist}
\end{figure}

For an initial assessment of the agreement between the classifications made by different inspectors, we compared the difference in the tail angles measured between different pairs of classifiers. In general, we find that the classifiers tend to agree remarkably well with each other, with the majority of the measures agreeing within a margin of 45 degrees. This is shown in Figure~\ref{fig:std_hist} where the scatter on the measured tail angles is shown for galaxies where at least 2 classifiers saw a tail (i.e. tail confidence level greater than $0$). 
The percentage of disagreement (with difference greater than 45 degrees) between pairs of classifiers averages at $\sim 12\%$ (ranging from $5\%$ to $23\%$). Interestingly, when comparing the confidence level assigned by two different classifiers to galaxies, we find that only in $43\%$ of the cases the pairs of classifiers agree on the confidence of the tail (i.e. no tail, marginal tail and clear tail), highlighting the subjectivity of this particular exercise. 

\begin{figure}
    \centering
    \includegraphics[width=0.48\textwidth]{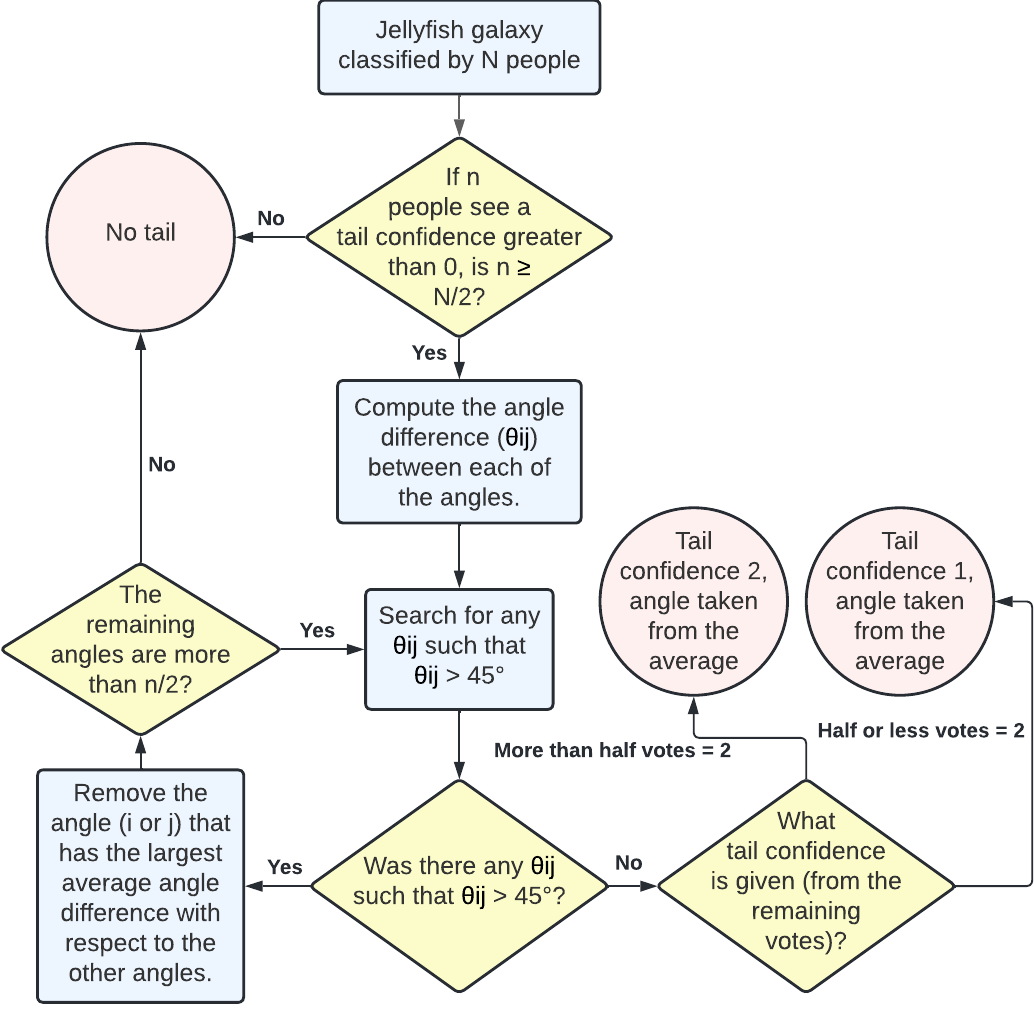}
    \caption{Flowchart followed to determine the mean tail angles of the stripped galaxies, and the associated uncertainty. This is computed by combining individual measurements of the angles ($\theta_i$) by $n$ classifiers and their declared confidence on their measurement.}
    \label{fig:flowchart}
\end{figure}

The final tail angle for each galaxy is  computed as the circular average from the results of all classifiers that agreed in the direction of the tail with a difference no greater than 45 degrees (i.e. rejecting outliers). The 45 degree threshold was chosen to be slightly larger than the average scatter obtained in the tail angle measurement (see solid red and dotted black lines in Figure~\ref{fig:std_hist}). 
The process of obtaining an average tail angle followed the flowchart from Figure \ref{fig:flowchart}, which is based on the work by \citet[][modifications were made to the angle rejection criteria to better suit a larger number of classifiers]{Kolcu2022}. This flowchart guarantees that the majority of the classifiers need to agree on the presence of a tail (confidence level greater than 0), and from that majority, there also needs to be a majority agreement on the tail direction. If these two condition are not met, the galaxy is classified as a jellyfish candidate with no tail. The final confidence level assigned to galaxies with tails is taken as the one with the higher number of votes. If a galaxy has an equal number of votes for marginal or clear tail, then we assign a marginal level for that galaxy. 

\begin{figure}
    \centering
    \includegraphics[width=0.48\textwidth]{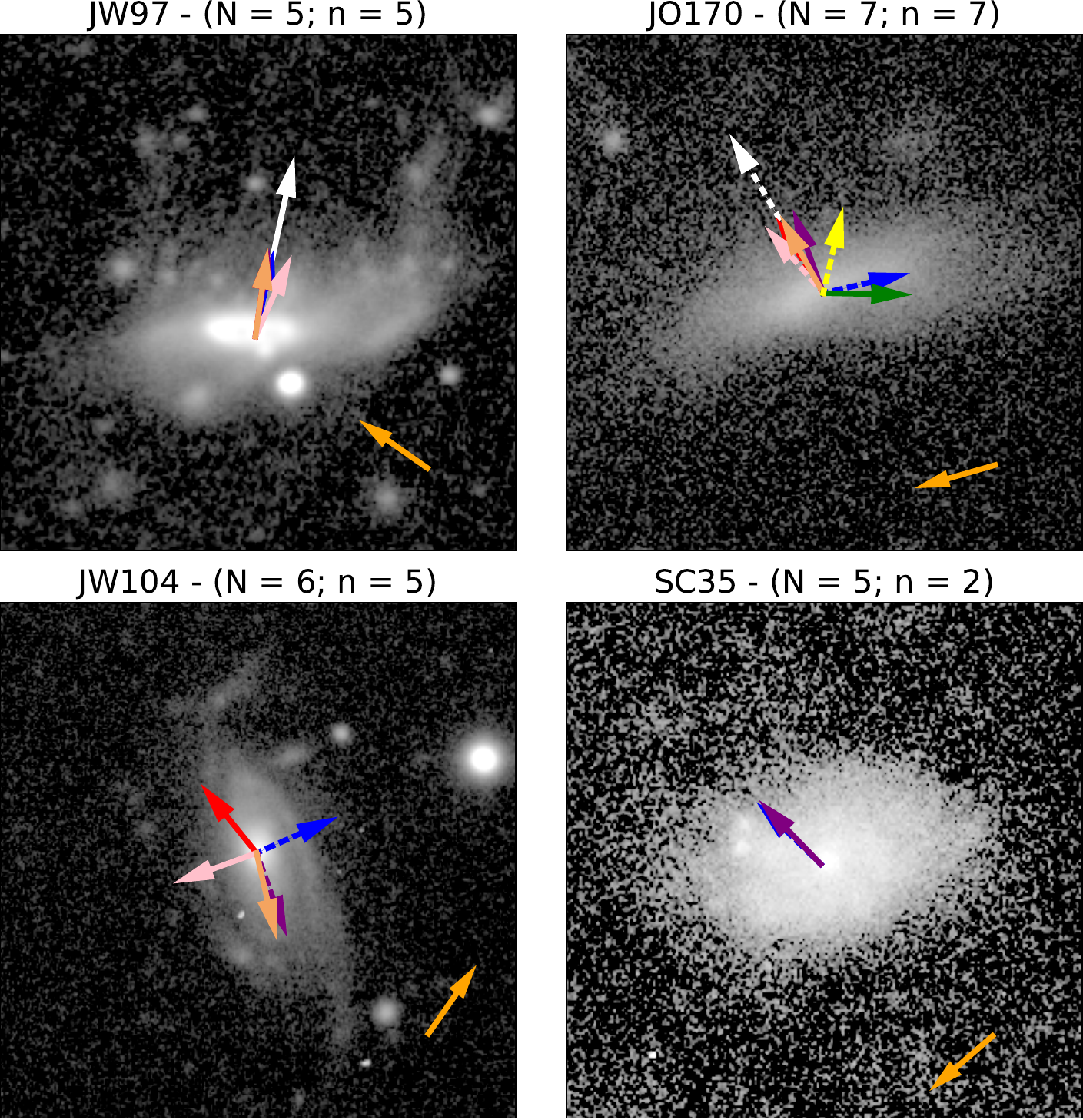}
    \caption{B-band optical images of 4 example jellyfish candidates for which tail angles were measured. Colored arrows represent the tail measurements of individual classifiers, while the larger white arrow shows the resulting mean tail, computed following the flowchart shown in Figure \ref{fig:flowchart}. Clear tails are represented with a solid line and marginal tails with a dotted line. The off-center orange arrows at the bottom right corner points to the direction towards the BCG of the cluster. The top of each image displays the galaxy name, the number of classifers ($N$), and the number of classifiers that see a tail ($n$).}
    \label{fig:CodeEx2}
\end{figure}

In Figure \ref{fig:CodeEx2} we present 4 randomly-selected examples. The tail direction measurements of the different classifiers are shown as colorful arrows, and the average tail direction is plotted as a bigger white arrow. We further checked the robustness of our tail measurements using broad-band optical images by comparing our measurements against the H$\alpha$ maps obtained from MUSE data for a sub-sample of 47 galaxies that were observed by GASP. This is shown in Appendix~\ref{ap:comparisson}, where we find that, although tails are often more clearly visible in H$\alpha$, there is a good agreement between the tail measurements performed in narrow H$\alpha$ vs. broad band optical images in 70\% of the cases. This test serves as support and provides
confidence to optical studies such as the present one.

\begin{figure}
    \centering
    \includegraphics[width=0.48\textwidth]{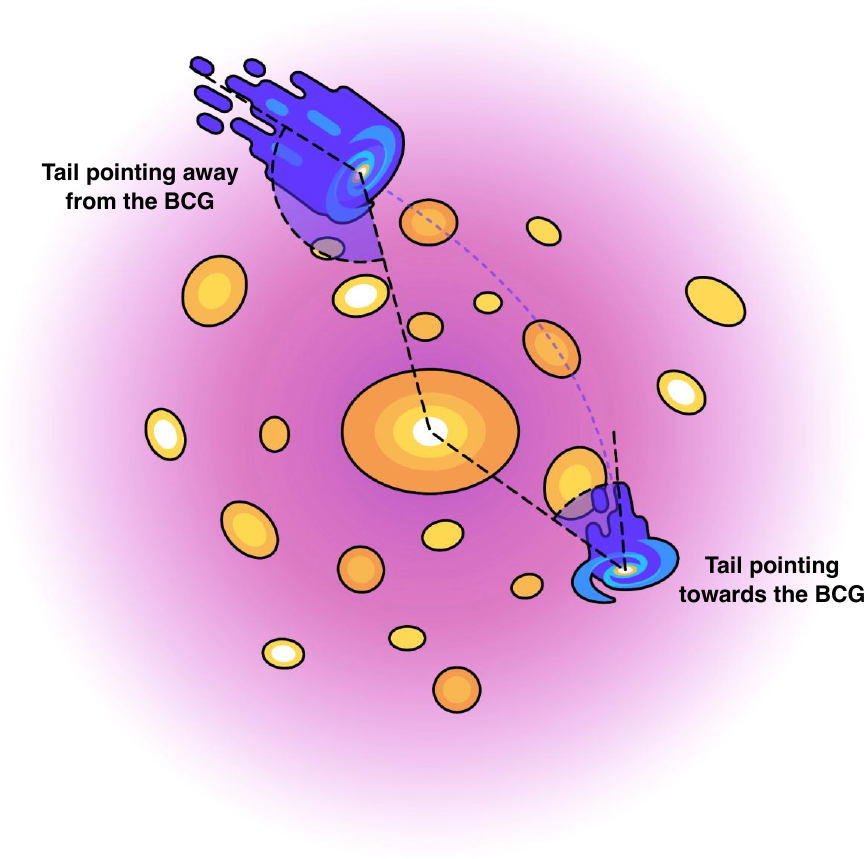}
    \caption{Illustration a jellyfish galaxy (blue) following a radial orbit (dotted blue line) in a galaxy cluster (orange ellipses) with a dense ICM (pink). The shaded blue angles represent $\theta$, which is the angle between the stripped tail and the direction towards the BCG (central orange galaxy). This angle can range between $0$ to $180$ degrees. It is assumed that the tails point in the opposite direction of motion. For the example galaxy, the tail first points "away" from the cluster center (large angle) and, after pericenter (i.e. closest passage through the cluster core), it points "towards" the center (small angle). Image credit: J. Utreras, J. Crossett \& Y. Jaff\'e}
    \label{fig:angle_diagram}
\end{figure}

\begin{table*}
\begin{center}
\caption{\label{table:angle_table_small} Table of mean angles for each jellyfish candidate. The x-axis angle denotes the counterclockwise tail angle with respect to the x-axis (with the north pointing up and west to the right). The tail-BCG angle denotes the angle of the tail with respect to the direction to the BCG. The confidence can take the values $0$, $1$, or $2$; representing no tail, marginal tail, or clear tail, respectively. The complete version of this table is available in the online version of this paper.}
\begin{tabular}{ccccccc} 
\hline
Galaxy & Cluster & \multicolumn{1}{c}{RA} & \multicolumn{1}{c}{DEC} & x-axis angle  & Tail-BCG angle & Confidence \\ & & \multicolumn{1}{c}{(deg)} &  \multicolumn{1}{c}{(deg)} &  \multicolumn{1}{c}{(deg)} &  \multicolumn{1}{c}{(deg)} & \\
\hline
JO1 & A1069 & 160.433 & -8.42 & 94.5 & 122.5 & 1 \\
JO2 & A1069 & 160.109 & -8.266 & 29.5 & 96.5 & 1 \\
JO3 & A1069 & 160.147 & -8.463 & 61.6 & 107.6 & 2 \\
JO4 & A1069 & 159.973 & -8.907 & 136.5 & 57.5 & 1 \\
JO5 & A1069 & 160.335 & -8.896 & 106.7 & 78.7 & 1 \\
JO6 & A119 & 14.242 & -1.299 & - & - & 0 \\
JO7 & A119 & 13.807 & -1.076 & - & - & 0 \\
JO8 & A119 & 14.487 & -1.336 & - & - & 0 \\
JO9 & A119 & 13.909 & -1.28 & 20.4 & 150.6 & 1 \\
JO10 & A119 & 14.423 & -1.312 & - & - & 0 \\
\end{tabular}
\end{center}
\end{table*}

After computing the mean tail angles with respect to the x-axis we transformed them into angles of the tails relative to the direction of the BCG, following the same convention used by \citetalias{Smith2022} and \cite{Kolcu2022}, such that angles close to $0$ degrees correspond to tails pointing towards the BCG, while angles close to $180$ degrees,  correspond to tails pointing away from the BCG, as shown schematically in Figure \ref{fig:angle_diagram}. In this work, we will refer to this angle as the tail-BCG angle. Table \ref{table:angle_table_small} provides the tail angle results for a subsample of jellyfish candidates. The complete table is available in the online version of the paper. 

From the tail classifications, we find that $71\%$ of the jellyfish candidates have tails, and of those, clear tails are found only in $39\%$ 
of the cases.
 
\section{Tail-BCG angle distribution}
\label{sec:tail-dist}

\begin{figure}
    \centering
    \includegraphics[width=0.5\textwidth]{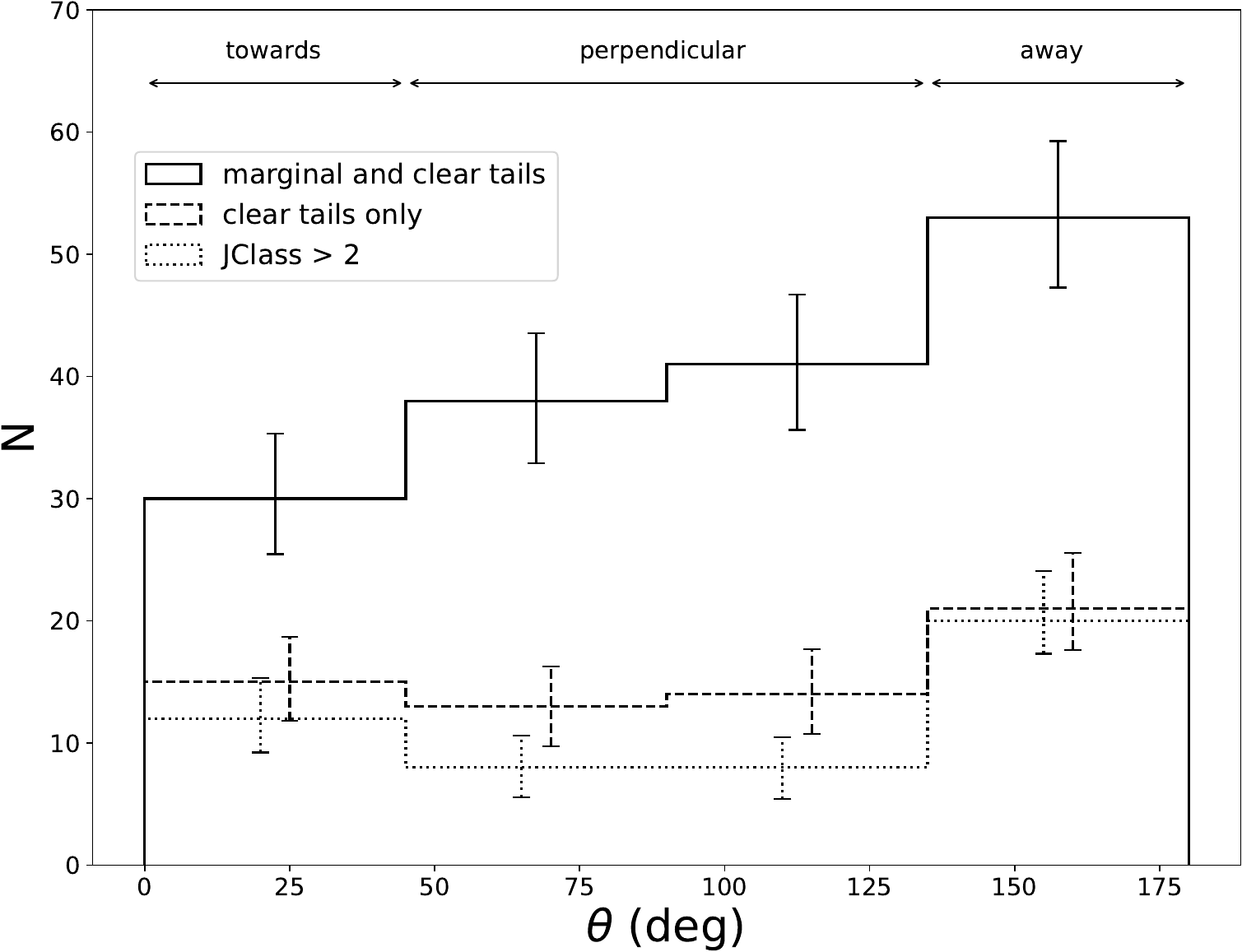}
    \caption{Tail-BCG angle distributions. All histograms are excluding interacting galaxies, galaxies within cluster mergers, and confirmed non-member galaxies. The histogram with a solid line represents all galaxies with tails ($162$ galaxies), the dashed line only includes galaxies with clear tails ($63$ galaxies), and the dotted line only includes galaxies with JClass greater than 2 ($48$ galaxies). Error bars were computed as the standard deviation from bootstrapping resampling.}
    \label{fig:tail_hist}
\end{figure}

Figure~\ref{fig:tail_hist}  shows the overall distribution of the jellyfish tail-BCG angles for the clean sample (i.e. excluding confirmed non-members, interacting clusters, and interacting galaxies). As marked in the figure, we define 3 categories of tails depending on their orientation:
\begin{itemize}
    \item "Towards" the cluster center, with tail-BCG angles $\theta < 45^{\circ}$. 
    \item "Perpendicular" to the cluster center, with $45^{\circ} \leq\theta < 135^{\circ}$.
    \item "Away" from the cluster center, with $\theta \geq 135^{\circ}$.
\end{itemize}
Note that the notation used in these classifications refers to the tail direction relative to the cluster center (defined by the BCG), and this is expected to be opposite to the direction of motion.

Using these definitions and considering the clean sample, the solid line in Figure~\ref{fig:tail_hist}  reveals a clear preference for galaxies to have tails pointing away from the cluster. More specifically, $32.7\%$ of the galaxies in the sample have tails pointing "away", while only $18.5\%$ are pointing "towards", and $48.8\%$ "perpendicular" (note that the perpendicular fraction reduces to $24.4\%$ if we consider that this sample spans twice the angle-bin size of the other samples). As will be shown and discussed in Sections~\ref{sec:orbits} and ~\ref{sec:discussion}, the observed tail-BCG angle distribution with a predominance of "away" tails is characteristic of a population of mostly infalling galaxies on radial orbits.

As mentioned above, galaxies in merging clusters were excluded from Figure~\ref{fig:tail_hist}  (along with interacting and non-member galaxies) to have the cleanest possible tail-BCG angle distribution. In Appendix~\ref{ap:merging-cl} we inspect the distribution of tail-BCG angles within interacting clusters only, and find that indeed it is different (much flatter) than the one found in regular systems. This finding justifies the exclusion of these clusters in our analysis, and support the notion that galaxies within unrelaxed clusters might be subject to particular conditions that alter the orbits of the galaxies and/or the medium surrounding them, which could have an effect on both the effectiveness of RPS and in the direction of the tails.

In Figure~\ref{fig:tail_hist} we further separate galaxies with JClass $> 2$ (dotted histogram) and galaxies with clear tails only (dashed), which represent the strongest and most confident cases of RPS. For these more confident RPS cases there is still a primary peak for galaxies pointing away from the cluster, but there is also a hint of  a secondary peak at low angles (tails pointing towards the cluster), together with a visible dip in the number of perpendicular tails (intermediate bins) not seen in the overall population (solid line). We performed a Hartigan's dip test \citep{HartiganDipTest} for unimodality on both subsamples to determine the significance of the apparent two-peaked distributions. For the JClass $> 2$ distribution we find a p-value of $0.020$, confirming the presence of a secondary peak. For the clear tails distribution, however, the peak and/or central dip is not significant enough.

\begin{figure}
    \centering
    \includegraphics[width=0.5\textwidth]{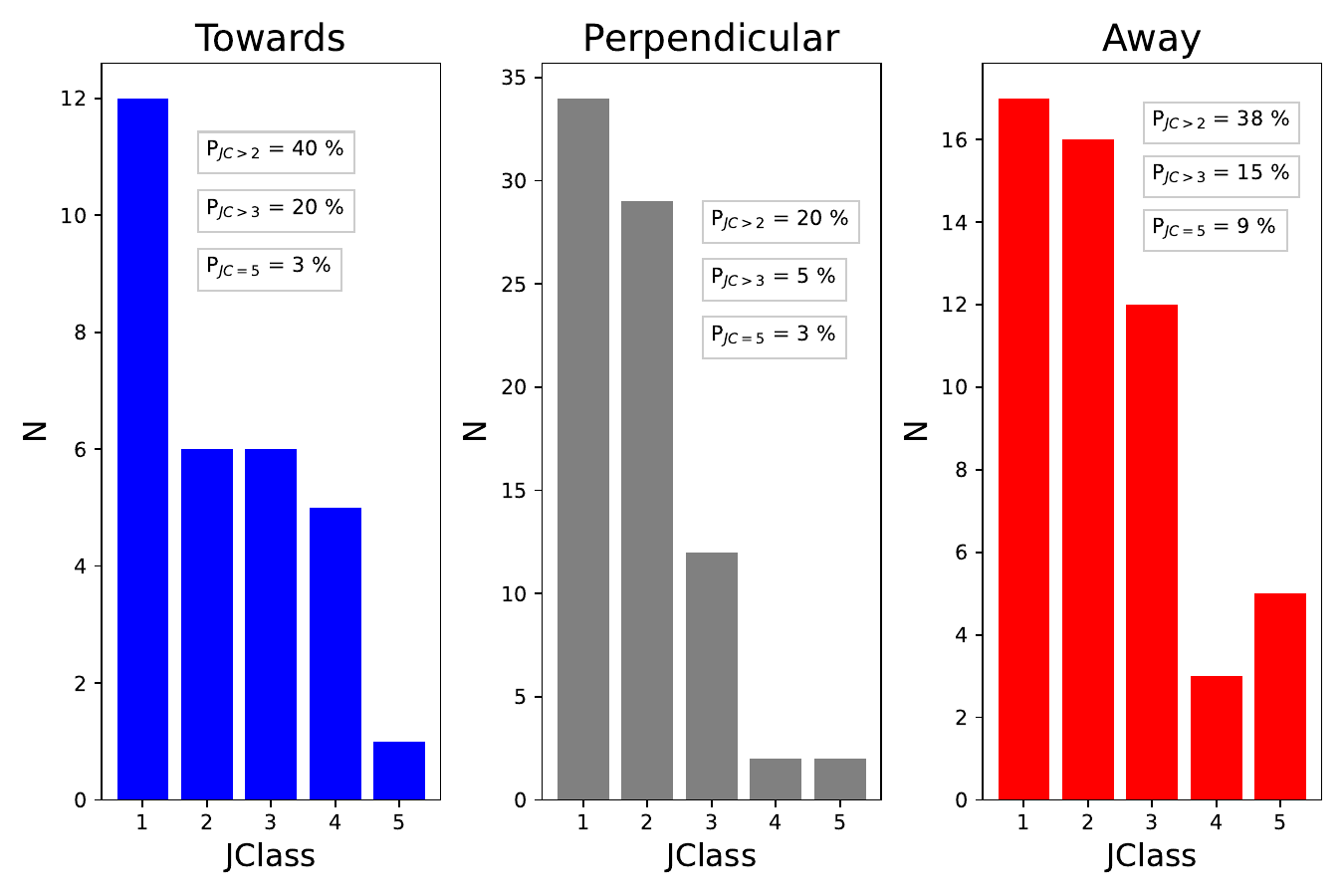}
    \caption{Number of galaxies on each JClass, for each tail-BCG angle sample from the histogram in Figure \ref{fig:tail_hist}. JClass goes from extreme cases (JClass 5) to progressively weaker cases, down to JClass 1. The distributions are excluding confirmed non-members, interacting galaxies, and interacting clusters.}
    \label{fig:jclass_hist}
\end{figure}

To inspect in more detail the strength of the RPS features for different tail orientations, in Figure \ref{fig:jclass_hist} we present the JClass distributions for the subsamples of tails pointing away, towards, and perpendicular to the cluster center. We find that galaxies in the towards and away samples have a higher relative fraction of galaxies with high JClass. For instance, only $5\%$ of the perpendicular tails have JClass $>3$, while $20\%$ and $15\%$ have JClass $>3$ in the towards and away samples, respectively. We also find that the highest fraction of JClass$=5$ galaxies (the most spectacular RPS candidates) is in the "away" sample. 

\section{Distribution of tails within the cluster}
\label{radial_distr}

In this section we present the position (Sec.~\ref{subsec:radial}) and velocity (Sec.~\ref{subsec:pps}) distribution of the tailed galaxies within the clusters, which will be later used (in Section~\ref{sec:orbits}) to constrain the lifespan of optical tails in RPS galaxies.

\subsection{Radial distribution}
\label{subsec:radial}

\begin{figure*}
    \centering
    \includegraphics[width=0.48\textwidth]{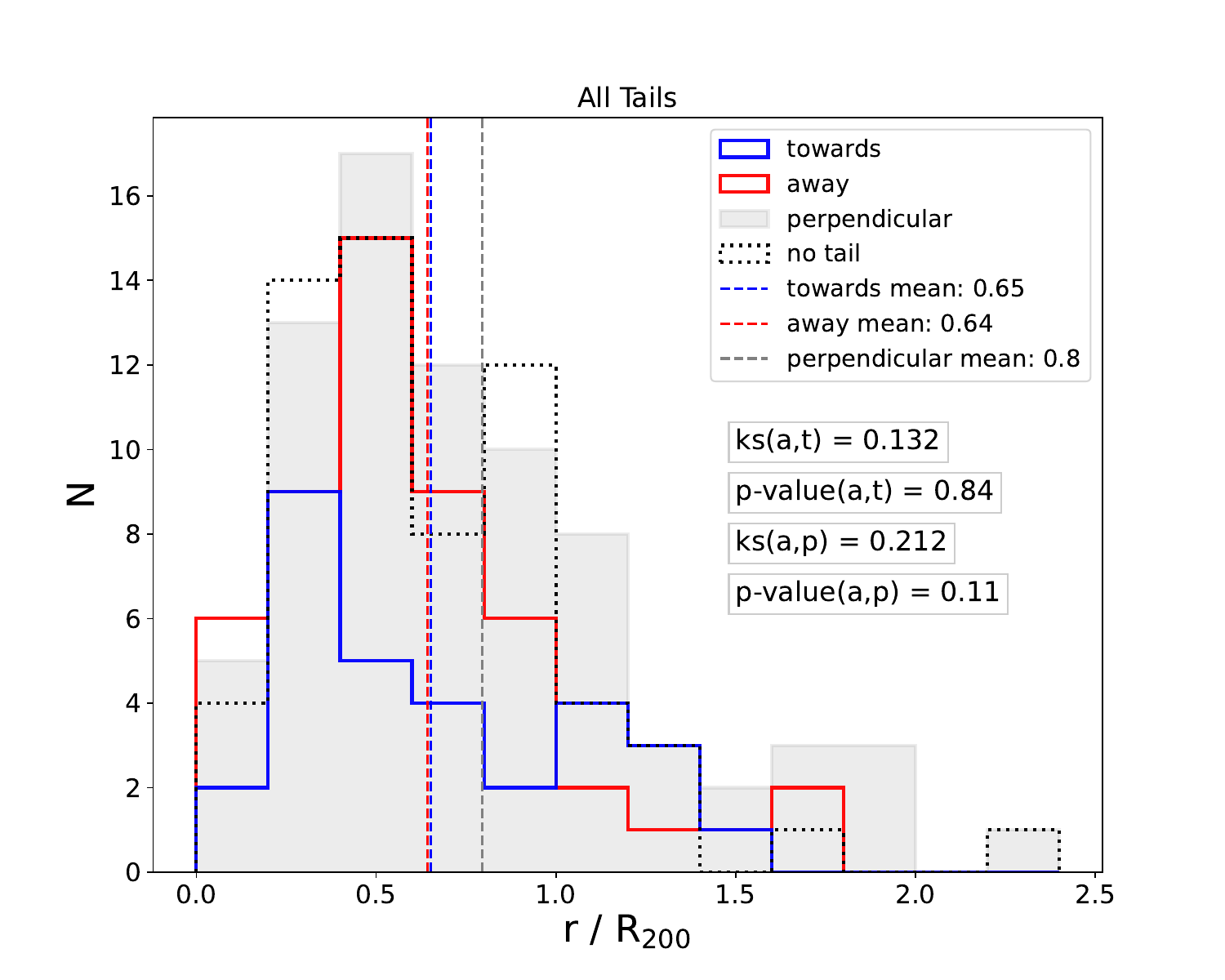}
    \includegraphics[width=0.48\textwidth]{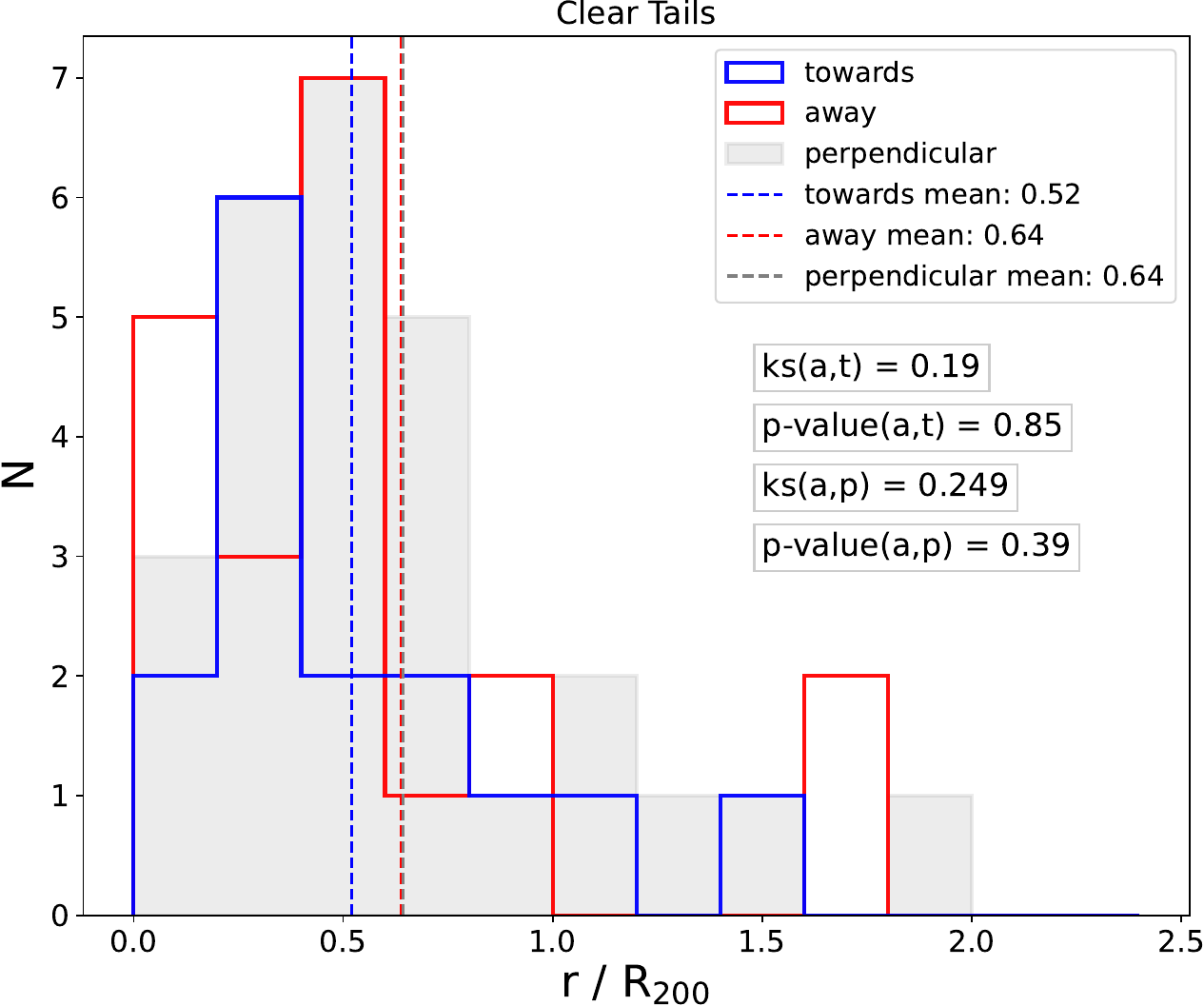}
    \caption{Projected radial distribution of jellyfish candidates with no tail (dotted), and tails pointing towards (blue), away (red), and perpendicular (gray) to the BCG. Left: Radial distribution for all tails. Right: Radial distribution for clear tails only. Colored dashed vertical lines represent the mean of the respective distributions. All plots include the results of the Kolmorov-Smirnov test, showing the KS statistic (ks) and p-values when comparing the distributions of the tails pointing away (a) with respect to the tails pointing towards (t) and perpendicular (p) to the BCG. The distributions are excluding confirmed non-members, interacting galaxies, and interacting clusters.}
    \label{fig:radial}
\end{figure*}

The left panel of Figure \ref{fig:radial} shows the radial distributions of all jellyfish candidates considered in this study (regardless of availability of spectra). The right panel shows the same but for clear tails only. 
We find that overall, the clear tail distribution is more centrally concentrated, consistent with the idea of stronger RPS nearer to the cluster center.

When splitting the sample by tail direction we find that the away sample peaks around $\sim0.5$ R$_{200}$, averaging at $0.64$ R$_{200}$ regardless of tail confidence. The other tail orientations show similar distributions, where the only noteworthy differences are a wider spread in the perpendicular tails, and a peak slightly closer to the center in the case of tails pointing towards. We note however that a Kolmogorov–Smirnov (KS) test \citep[see][]{KS_test_related_paper} suggests that the radial distribution of the away sample is not significantly different than that of the other tail directions. 

\begin{figure*}
    \centering
    \includegraphics[width=0.48\textwidth]{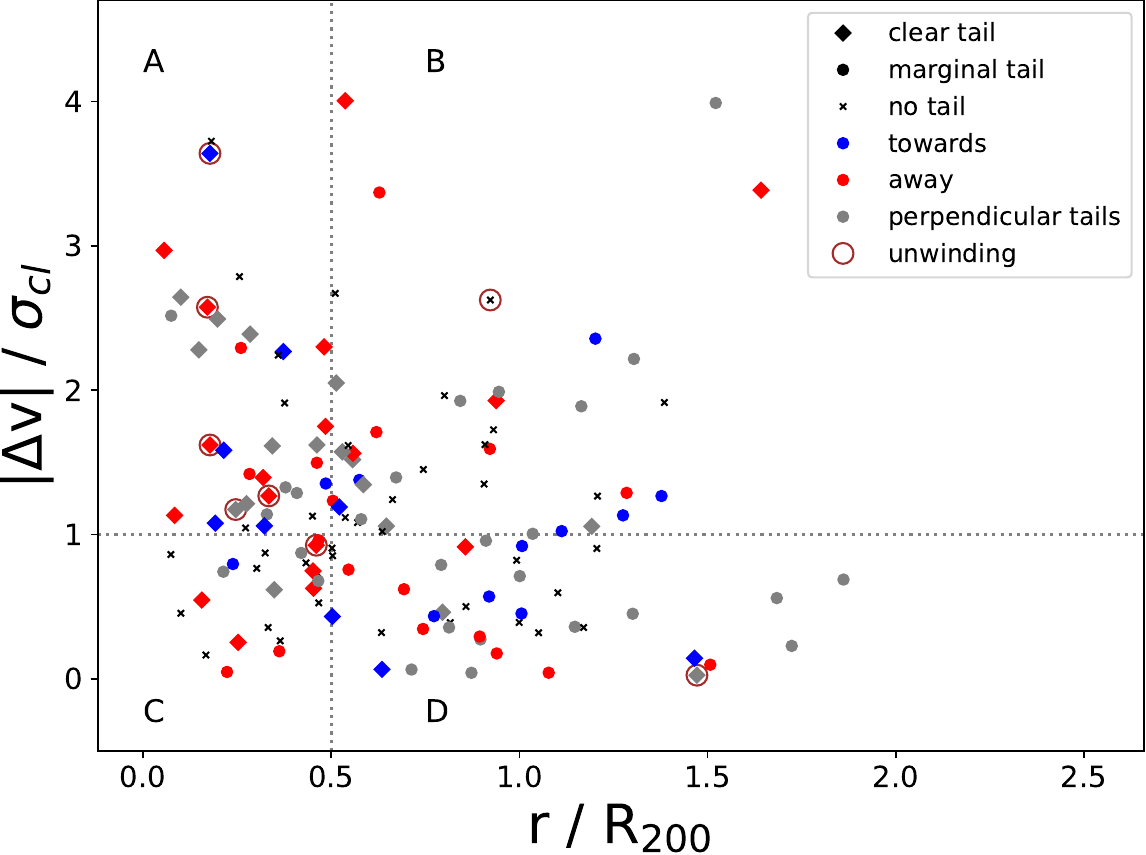}
    \includegraphics[width=0.48\textwidth]{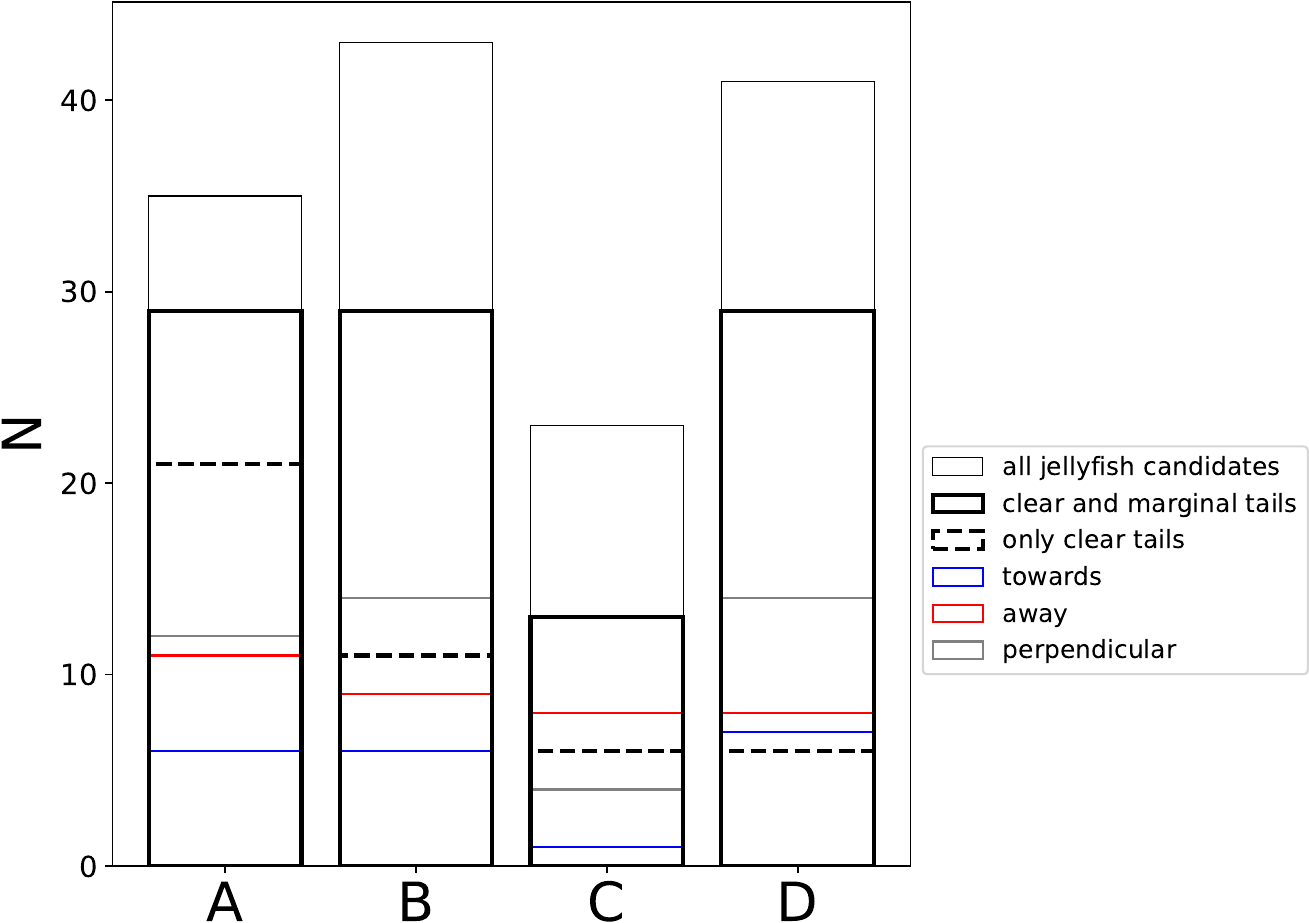}
    \caption{Left: Projected phase-space diagram for spectroscopically confirmed members with stripping signatures, excluding gravitationally interacting galaxies, or galaxies in interacting clusters. We highlight tails pointing towards (blue) and away (red) from the BCG, as well as those with perpendicular (gray) tails and no tails (black). Dotted lines divide the phase-space diagram at $r/R_{200}=0.5$ and $\Delta v/\sigma = 1$, into four regions: A, B, C, D. Note that the line of sight velocity is shown in absolute value. Right: Number of jellyfish candidates on each region (A, B, C, D) from the phase-space diagram. Bars with a narrower solid line represent candidates with any tail confidence (including no tail), while the bars with thicker lines only include candidates with tails. Dashed bars only include galaxies with clear tails. Colored bars represent the number of galaxies of different tail orientations (colored as in the left panel).}
    \label{fig:PPS}
\end{figure*}

\subsection{Phase-space distribution}
\label{subsec:pps}

To further trace the orbital histories of the galaxies we study the locations of jellyfish candidates in a projected position vs.  velocity phase-space diagram for those galaxies with available redshifts. In the left panel of Figure~\ref{fig:PPS} we show the phase-space diagram  for the subsample of $141$ confirmed members (from which $100$ have tails), distinguishing between tail confidence and tail orientations using different colors. 
The normalized line-of-sight velocities in the y-axis were computed as: 
\begin{equation}
    \centering
    \frac{\Delta v}{\sigma_{\text{cl}}} = \frac{c\left(z-z_{\text{cl}}\right)}{\left(1+z_{\text{cl}}\right)\sigma_{\text{cl}}} \text{,}
    \label{eq:vlos}
\end{equation}

\begin{figure}
    \centering
    \includegraphics[width=0.48\textwidth]{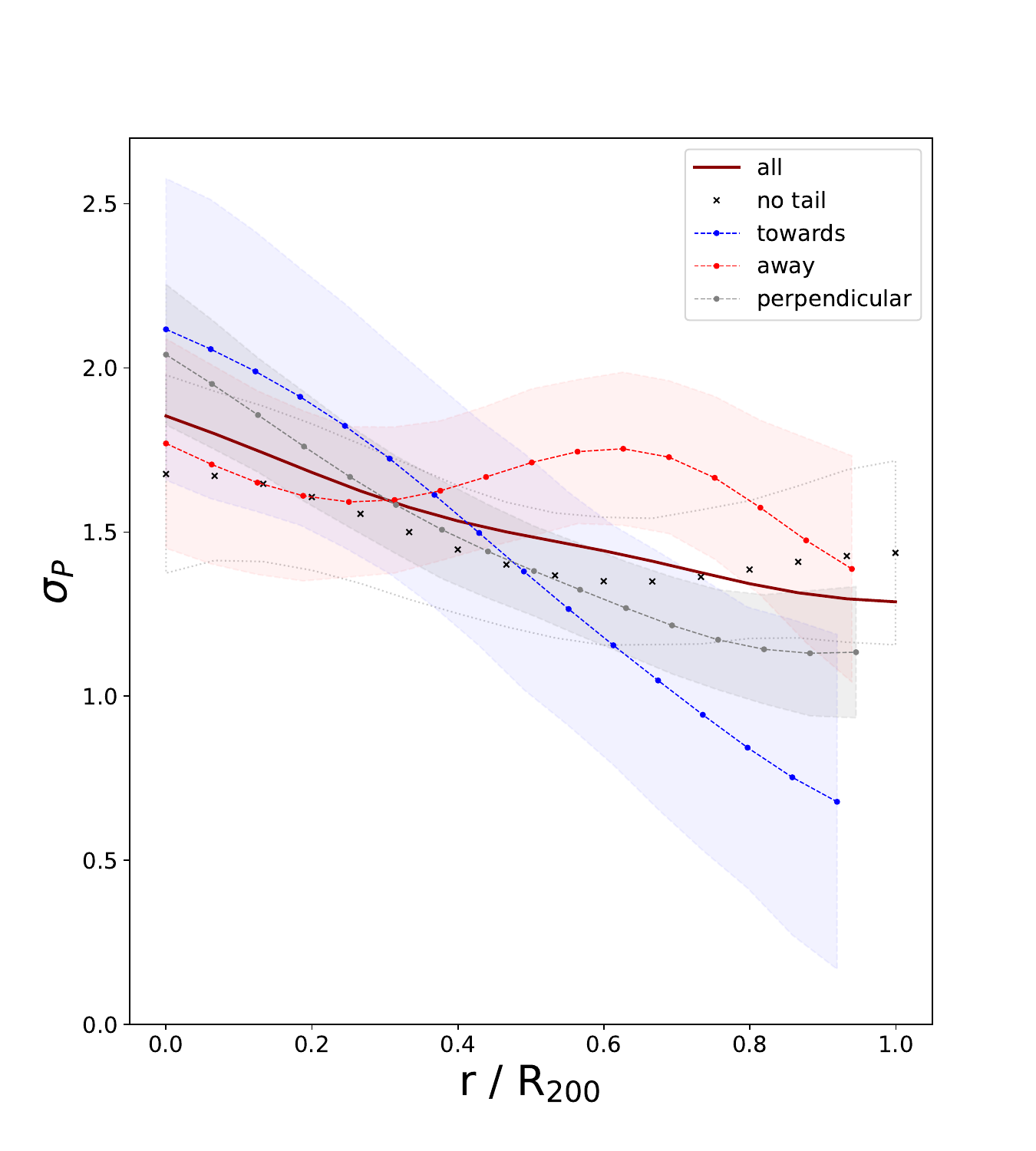}
    \caption{Normalized line of sight velocity dispersion radial profiles of the galaxies on the phase-space diagram in the left panel of Figure \ref{fig:PPS} (up to $1$ R$_{200}$). We show the radial profile for the subsamples of galaxies with no tail (black "x" markers), galaxies with tails pointing towards (blue dashed line), away (red dashed line), and perpendicular (gray dashed line) to the BCG. The radial profile for all confirmed members (dark red solid line) is added for reference. The velocity dispersion is computed following the method described in \protect\cite{Bilton2018}. Error bands represent the standard deviation of 1000 Monte Carlo resamples. We do not show the error band for the dark red solid line to not overcrowd the plot, but we note it has an average value of $\pm0.12$.}
    \label{fig:VDP}
\end{figure}

We divide the phase-space diagram in regions of interest in a similar way to \citetalias{Smith2022}:  with boundaries at $r/R_{200}=0.5$ and $\Delta v/\sigma = 1$, defining four regions (labeled A, B, C, D in Figure \ref{fig:PPS}). In the right panel of Figure \ref{fig:PPS} we present the jellyfish candidate counts on each region. In the upper left region (A) of phase-space, where we expect RPS to be the strongest \citep[see][]{Jaffe2018}, we find that $72\%$ of the galaxies with tails have a high confidence classification. This represents the highest fraction of clear tails. The second highest fraction of clear tails is in the lower left region (C) with $46\%$, followed by the upper right region (B, $38\%$) and the lower right region (D, $21\%$). This is consistent with ram pressure starting when the galaxies enter the cluster (at high radii and low velocities), and developing stronger signatures of stripping as they approach the cluster core, especially those with high velocities. 

When splitting the sample by tail direction we find that galaxies with tails pointing away from the cluster are in all regions of phase-space, but have a mild preference for the high velocity regions, consistent with an infalling population. Galaxies with towards tails could potentially be associated with cases past pericenter and they are mostly found in the regions A,B,D, with very few examples in region C. However, we note that the towards cases at high distances tend to have less clear tails. The same is true for perpendicular cases which concentrate on quadrants A,B,D.  

To complement the phase-space diagram we also constructed a velocity dispersion profile (VDP) of the jellyfish candidates in the phase-space sample. We calculate the velocity dispersion using the method first introduced by \cite{Bergond2006} for globular clusters, corrected and adapted to galaxy clusters as in \cite{Bilton2018}. This method calculates the velocity dispersion in radial bins using a exponentially weighted Gaussian window function $\omega_{i}$, given in equation 2 from \cite{Bilton2018}, such that the VDP is described by

\begin{equation}
    \centering
    \sigma_{\text{P}} \left(r\right) = \sqrt{\frac{\sum \omega_{i} \left(\Delta v_{i}\sigma_{\text{cl},i}^{-1}\right)^{2}}{\sum \omega_{i}}} \text{,}
    \label{eq:VDPeq}
\end{equation}

where $\Delta v_{i}\sigma_{\text{cl},i}^{-1}$ is the normalized line of sight velocity of each galaxy inputted, computed as in Equation \ref{eq:vlos}. 

In Figure~\ref{fig:VDP} we show the VDP for each tail orientation. Note that error bands are large due to the low number statistics of the sample, which should be taken into consideration when analyzing this result. For this reason, we consider in our analysis galaxies up until $1$ R$_{\text{200}}$ since there are too few examples of each tail orientation farther than that.

From Figure~\ref{fig:VDP} we find that the VDP of jellyfish candidates with tails pointing away from the cluster are generally higher when compared with the perpendicular and towards tails, at least outside the cluster core. In the inner parts of the clusters we do not observe significantly different VDPs between different tail orientations. 

For galaxies with tails pointing towards the cluster center, we see a steep decrease in the VDP from small to large clustercentric distances. Figure~\ref{fig:VDP} also shows that tails pointing towards  have the largest velocity dispersion near the center, albeit with large uncertainty. This result is largely driven by the galaxy JO201 at the top left of the phase-space diagram, which is an extreme case of stripping along the line of sight in a moment close to pericentric passage \citep{Bellhouse2017}.

The VDP of the jellyfish candidates with no tail (black solid line in Figure~\ref{fig:VDP}) has the lowest velocity dispersion near the cluster center (although not significantly different from the other galaxies within errors), which is consistent with less intense ram pressure.
Indeed  galaxies with no tail were identified as jellyfish candidates but are likely milder or less clear cases of stripping. However, at distances larger than $\sim0.5$ R$_{\text{200}}$ they have the second largest velocity dispersion.

\begin{figure*}
    \centering
    \includegraphics[width=0.97\textwidth]{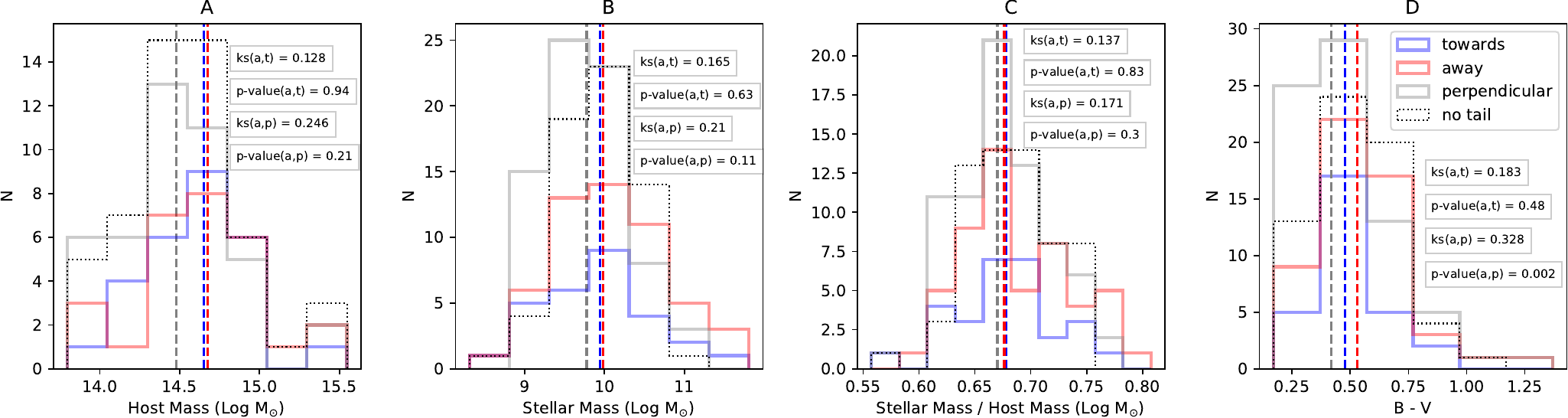}
    \caption{Distributions of cluster and galaxy properties for the populations of galaxies without tails (dotted), and tails pointing towards (blue), away (red), and perpendicular (gray) to the BCG. Dashed vertical lines represent the median of the samples for the blue, red, and gray histograms. All plots include the results of the Kolmorov-Smirnov test and p-values when comparing the distributions of the tails pointing away (a) with respect to the tails pointing towards (t) and perpendicular (p) to the BCG. A: Host mass (M$_{200}$) distribution of clusters hosting any galaxy from a given population. B: Stellar mass distributions of the galaxies. C: Stellar to host mass ratio distribution of the galaxies. D: Galaxy color distribution. All plots exclude interacting galaxies, interacting clusters, and confirmed non-members.}
    \label{fig:combined_prop_hists}
\end{figure*}

\section{Dependence of tail-BCG angle on cluster and galaxy properties}

\citetalias{Smith2022} found that, for radio continuum stripped candidates, the away sample prefers higher mass hosts, lower mass galaxies, and lower mass ratios. Here we explore in a similar way how our sample of optical jellyfish candidates varies with cluster mass, galaxy mass, mass ratio, and galaxy color.

\subsection{Dependence on cluster and galaxy mass}
\label{sec:mass}

The first (left) panel of Figure \ref{fig:combined_prop_hists} shows the cluster (host) mass distribution of clusters hosting any galaxy of the respective subsample (i.e. no tails, tails pointing towards,  away or perpendicular to the cluster center)\footnote{Note that the same cluster can be assigned to more than one subsample if it hosts galaxies of different tail orientations}. We find that the away and towards samples are similar, but the perpendicular tails seem to inhabit lower mass clusters,  although statistically (KS test) the cluster mass distributions of all samples are not significantly different. 

In the second panel of Figure \ref{fig:combined_prop_hists} the  stellar mass distributions for the different subsamples are plotted.Again we find no significant differences. The same occurs on the third panel where we plot the distribution of the ratio between the stellar mass and the host mass, where the differences are negligible. 

To check for potential correlation (and hence bias) between the tail angles  and stellar mass, we attempted to compare the tail-BCG angle distribution of the most massive (M$_{*}> 10^{10} M_{\odot}$) galaxies with those having lower-masses (M$_{*}< 10^{10} M_{\odot}$; not shown), and found that the high mass galaxies have a more distinct peak at high angles with other angles having equally low counts, while the low mass sample has a gradually increasing distribution with angles, as the one seen in Figure~\ref{fig:tail_hist} (black line). However, the number statistics are low and a KS test between both subsamples suggests the distributions are not too dissimilar (p-value of $0.23$). Because the number statistics are low. We leave a full exploration of the effect of stellar mass on tail directions and stripping timescales for a future study.

In summary, we do not find significant mass segregation when considering different tail orientations, maybe only with the exception of perpendicular tails (but not significant according to the KS test), which appear to prefer slightly lower mass clusters and lower galaxy stellar masses.

\subsection{Dependence on color}

\begin{figure}
    \centering
    \includegraphics[width=0.48\textwidth]{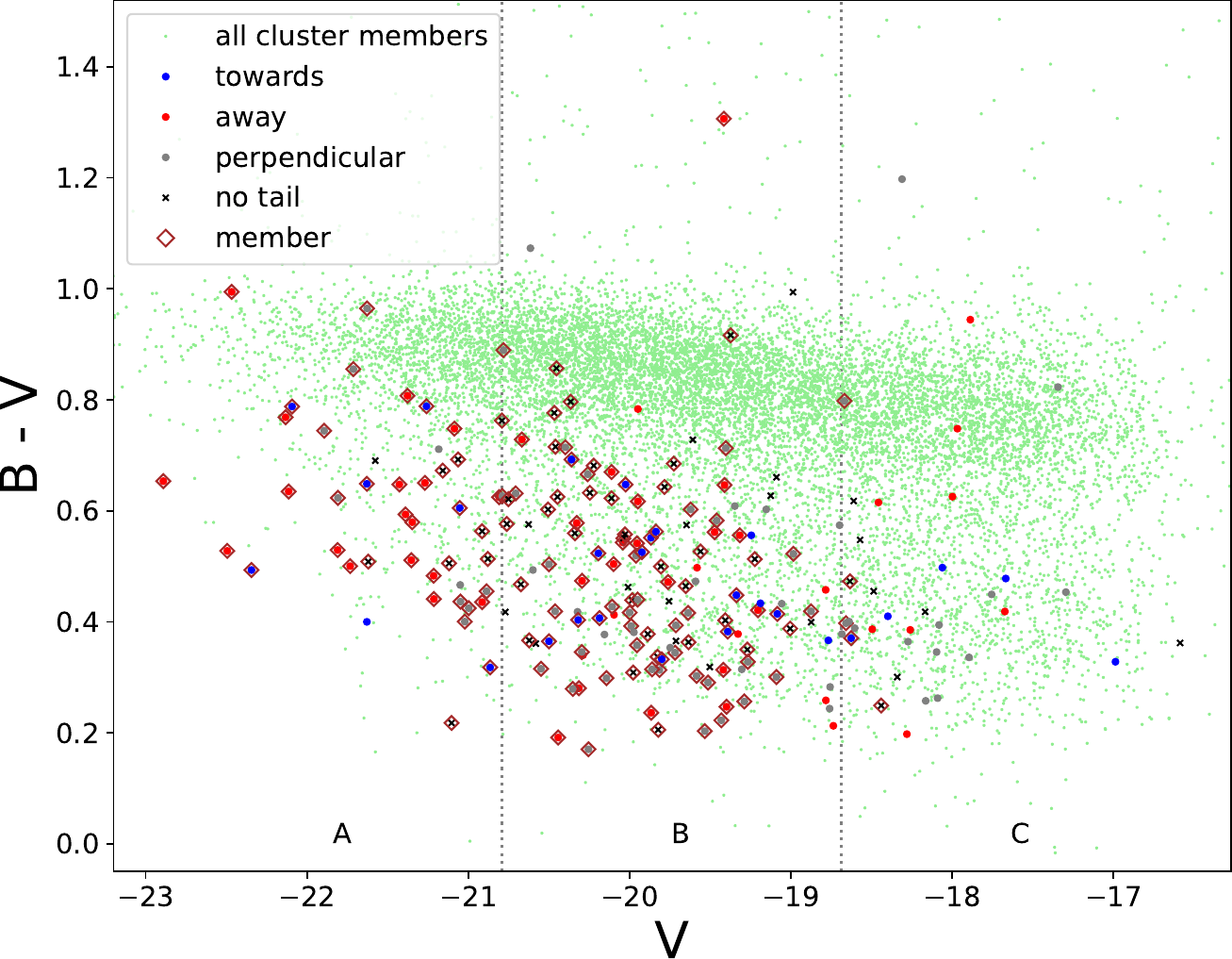}
    \caption{Color-Magnitude Diagram of the jellyfish candidates. The tail orientation of the galaxies is highlighted with colors; towards (blue), away (red), perpendicular (gray), and no tail (black). Confirmed members are highlighted with brown diamonds. The green points represent the colors of all the cluster members from OmegaWINGS and WINGS (including non-jellyfish candidates). We divided the diagram into three magnitude regions A, B, and C, going from brighter to fainter, with a width of $\sim{2.1}$ V-mag. This plot excludes interacting galaxies, interacting clusters, and confirmed non-members.}
    \label{fig:CMD}
\end{figure}

We further study jellyfish tail directions as a function of color, to test (indirectly) whether the orbital history of the galaxy could be reflected in its stellar populations. 
In Figure \ref{fig:combined_prop_hists} we see a slightly greater (and statistically significant) difference in color between the subsamples than in the mass comparisons. Because of this, we inspect in more detail how the measured tail-BCG angles depend on galaxy color, which is a broad indicator of the age of the stellar populations of the galaxies (although metallicity would also have some influence in color). In Figure~\ref{fig:CMD} we show the color-magnitude diagram (CMD) of the sample of jellyfish candidates and, for reference, the sample of WINGS and OmegaWINGS cluster galaxies in the background, where the red sequence of passive galaxies is clearly separated from the blue cloud of star-forming ones. Most jellyfish candidates belong to the blue cloud (mostly below the red sequence), as expected of gas-rich late-type galaxies that have not yet been completely stripped (not quenched).

From the last panel of Figure \ref{fig:combined_prop_hists} we find that perpendicular tails have the bluest colors. Furthermore, this is the only case where the KS test yields a low p-value ($0.002$), confirming that the colors of perpendicular tails follow a significantly different distribution from the other tail orientations.

Interestingly, the jellyfish candidates with tails pointing towards the cluster center (presumably post-pericentric passage) also have slightly bluer colors than the ones pointing away from the cluster (infalling). However, this color difference is not significant according to the KS test. Furthermore, if we split the CMD in Figure~\ref{fig:CMD} into three regions from brightest to faintest (left to right; A, B, C, respectively), we find that the galaxies from the towards sample in the faint end (region C) are the only ones shifted to bluer colors. When inspecting these galaxies we note that four have low JClass and low tail confidence. Only one galaxy has a clear tail and is the reddest of the five. Furthermore, most faint galaxies are non-confirmed members. Therefore, the slight difference in color between the away and towards samples is only caused by a small number of low-confidence measurements.

\begin{figure*}
    \centering
    \includegraphics[width=0.98\textwidth]{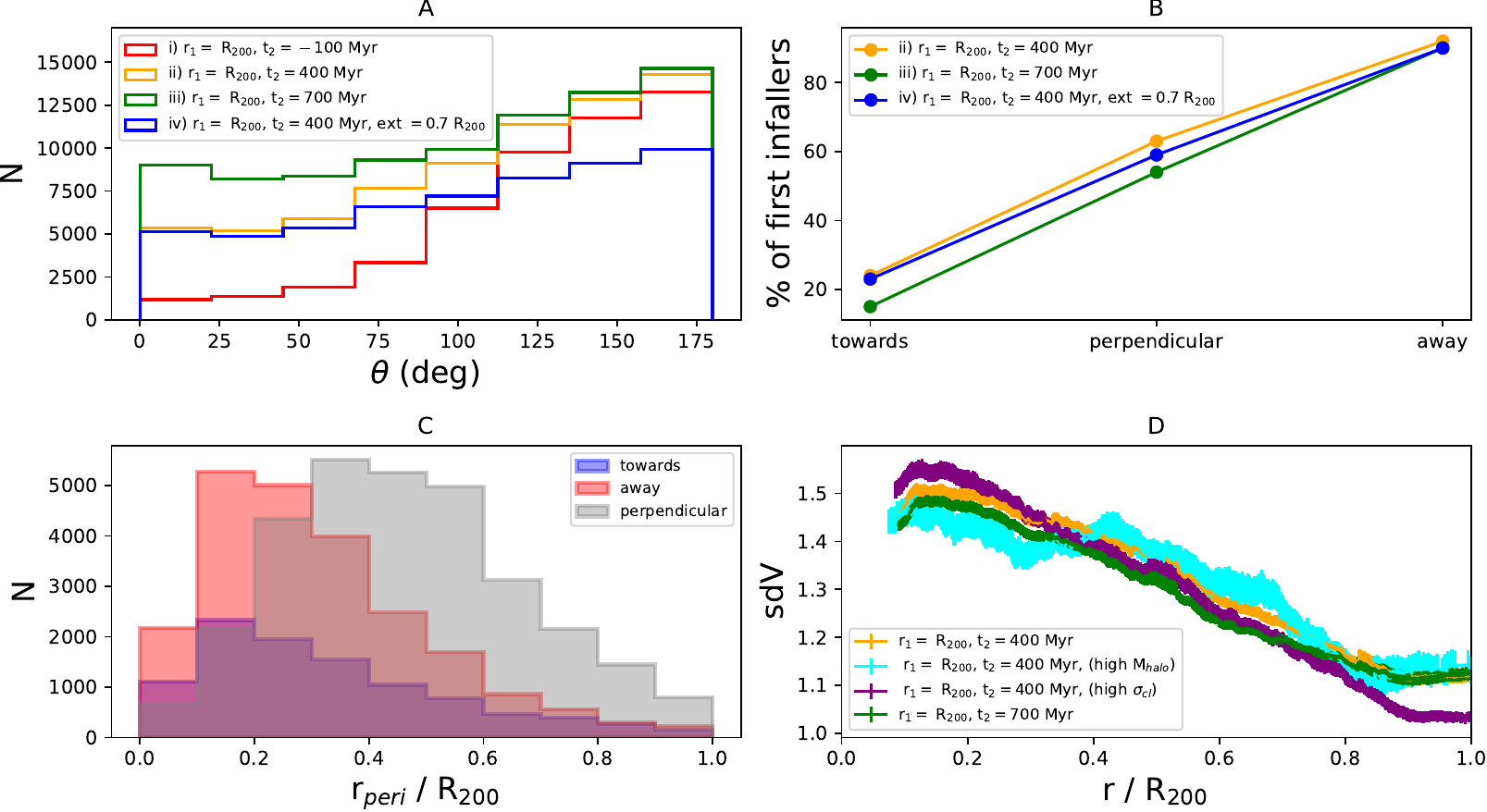}

    \caption{Simulated results. Panel \textbf{A)} Distribution of projected tail-BCG angles for the different simulations in Table \ref{table:models} (i, ii, iii, iv). \textbf{B)} Fraction of infalling galaxies (pre-pericentric passage) in the 3 tail-BCG angle bins considered in this paper (i.e. tails pointing "towards", "perpendicular" and "away" from the cluster. \textbf{C)} Pericenter distance distribution for the simulated samples (model ii) of tails pointing towards (blue), away (red), and perpendicular (gray) to the BCG. The tail-BCG angles used for this plot are projected angles. \textbf{D)} Fiducial VDP of the simulation (model ii, orange), for high cluster velocity dispersion (purple), for high halo mass galaxies  ($> 10^{12}$, cyan), and for high values of $t_{2}$ (model iii, green). The velocity dispersion (sdV) is computed as the standard deviation of the projected line of sight velocity.
   }
   \label{fig:simulated_before_after_peri}
\end{figure*}

\section{Orbits of jellyfish galaxies and lifespan of their tails}
\label{sec:orbits}

In order to use our tail direction results to  constrain the orbits of jellyfish galaxies and the lifespan of optical tails, we compare our results with models generated from simulation data following the method introduced by \citetalias{Smith2022}. In short, this method uses N-body cosmological dark matter only simulations, in which the galaxy tails are later added using a set of three free parameters; $r_{1}$, $\delta$, $t_{2}$. The parameter $r_{1}$ is the 3D distance from the cluster center at which the tails first become visible. The tail direction is expected to be opposite to the direction of motion of the galaxy. However, if the galaxy changes orbital direction, it takes some time for the tail to change direction (see \citeauthor{Roediger2007} \citeyear{Roediger2007}, \citeauthor{Tonnesen_2019} \citeyear{Tonnesen_2019}). To account for this the parameter $\delta$ is used to set the delay that takes for the tail to change direction when the galaxy has changed its orbital direction. Lastly, the parameter $t_{2}$ is the time after pericenter that it takes for the galaxies to lose their tail. If the galaxies lose their tails before pericenter, then $t_{2}$ can take negative values. 

\subsection{Simulated tail directions}
\label{mock_sims}

In this subsection, we consider 4 idealised model scenarios with preset fixed parameters, summarized in Table~\ref{table:models}. In the following subsection, we will attempt to constrain the parameters using the observations. But here, we systematically vary the time the tails remain visible after pericenter, in order to deepen our understanding of the results from the observational sample. In essence, model i represents a case where tails disappear before pericentric passage, models ii and iii  show cases where tails last for 400 and 700 Myr after pericenter, and model iv is similar to model ii but with limited radial data coverage (only considers projected clustercentric distances $< 0.7 R_{200}$, as is the case for some clusters in our sample). 

\begin{table}
\begin{center}
\caption{\label{table:models} Parameters of the 4 models considered.  The model parameters $r_{1}$ and $t_{2}$ are listed in columns 2 and 3. Negative values of $t_{2}$ correspond to tails disappearing before pericentric passage. Model iv) further considers a case where the data is limited to a a given projected clustercentric distance (extent).}
\begin{tabular}{cccc} 
\hline
Model & r1 & t2 & Extent  \\
\hline
i   & R$_{200}$ & -100 Myr & No cut          \\
ii  & R$_{200}$ & 400 Myr  & No cut          \\
iii & R$_{200}$ & 700 Myr  & No cut          \\
iv  & R$_{200}$ & 400 Myr  & $0.7$ R$_{200}$
\end{tabular}
\end{center}
\end{table}

In Figure \ref{fig:simulated_before_after_peri} we present a set of plots from the simulated data. In Panel A we show the projected tail-BCG angle distribution for the 4 models, which clearly show a steep monotonically increasing distribution for model i. This is expected for a population of galaxies that have not crossed pericenter yet, as objects with tail-BCG angles $<90$ degrees can only arise via projection effects. Models ii and iii on the other hand have an increased number of galaxies with lower tail-BCG angles, which correspond to the ones with tails still visible after crossing pericenter. Indeed, the presence of objects with tails that remain visible after pericenter is a requirement to explain the turn up in numbers in the lowest angle bin (seen in the observations; Figure~\ref{fig:tail_hist}). When restricting the coverage of the data (compare model iv to model ii) we find we systematically lose objects with large tail-BCG angles. This is because objects at large projected distances from the cluster are predominantly those with tails pointing away from the cluster. 

Panel B further shows the fraction of galaxies on first infall (i.e. before first pericentric passage) in the different models in 3 bins of tail-BCG angle: for galaxies with tails pointing towards, perpendicular, and away from the cluster. We find that in all cases the tails pointing away from the cluster are highly dominated by first infallers ($\sim 90 \%$). On the contrary, galaxies with tails pointing towards the cluster have a low fraction of first infallers, which confirms that these are mostly galaxies that have  passed pericenter. In fact Model iii (which allows tails to live the longest after pericenter) has the lowest fraction of first infallers in the "towards" bin, which further emphasizes this point. Finally, we find that perpendicular tails are a roughly equal mix of pre- and post-pericentric passage galaxies (with a slight preference for pre-pericenter) which suggests the population of galaxies with perpendicular tails is likely dominated by galaxies on less radial orbits. Furthermore, in Panel C we show the distribution of the distance between the pericenter of the orbits with respect to the cluster center, $r_{\rm peri}$, for the galaxies in each of the 3 tail bins, and its clearly visible that perpendicular tails have the largest $r_{\rm peri}$ of all tailed galaxies, confirming these are on less plunging orbits. 

Finally, we used the idealised models to examine the impact of various parameters on the shape of the VDP. The velocity dispersion is computed at each projected radius with a moving window of width 1$\%$ of the total particles. We bootstrap on the particle in the window 100 times, and the thickness of the VDP trend line denotes the one-sigma deviation between bootstraps. In general, we found the VDP was insensitive to many of the parameters we tested. However, in Panel D of Figure~\ref{fig:simulated_before_after_peri} we highlight some interesting examples of the model VDPs. Model ii is the orange line, and we compare it with the other models where one parameter is varied. For example, the cyan curve is for galaxies with a high dark matter halo mass (M$_{\text{halo}} > 10^{12} M_{\odot}$)\footnote{This limit roughly corresponds to a stellar mass of M$_{*}\sim10^{10.5}$ $M_{\odot}$  when using the M$_{*}$ vs M$_{\text{halo}}$ scaling relation from \cite{DiTeodoro2023}}.  In this case, the galaxies tend to have higher dispersion outside the cluster core (projected radius $>$0.4~R$_{200}$), but have reduced values in the inner region because of backsplash galaxies moving slower as a result of dynamical friction. Clusters with a high velocity dispersion ($>$550 km/s) have a steeper profile with a larger dispersion near the cluster center, which can be a result of this cut preferentially selecting clusters with high velocity galaxies near the cluster core. Finally, a slight change is observed when the $t_{2}$ parameter is increased to 700 Myr (model iii), as this causes more backsplash galaxies (whose orbital velocities are lower) to be seen, which in turn reduces the velocity dispersion over a broad range of projected radius.  

Overall, the simple models applied to the simulations show that the observed tail distribution is consistent with a population of galaxies with tails that appear during first infall and dissapear after the first pericenter passage and not before. The simulations also show that most galaxies with perpendicular tails indeed follow less plunging orbits when compared to the galaxies pointing towards or away from the cluster center. Finally we did not find significant variations in the velocity dispersion profiles of the different samples. 

\subsection{Bayesian Parameter Estimation results}
\label{MCMC}

\citetalias{Smith2022} performed a Bayesian parameter estimation to constrain the $r_1, t_2$ and $\delta$ parameters using radio continuum observations of jellyfish candidates from \cite{Roberts2021}. To do this, they produce simulated phase-space and tail orientation distributions and obtain the probability density functions (PDFs) of model parameters by sampling the posterior distribution using the Markov Chain Monte Carlo method. They had a large observational sample of galaxies covering up to R$_{180}$ ($\sim1.05$ R$_{200}$) of the cluster. Ideally, we would want a cluster sample reaching much farther than one virial radius, covering up to the infall regions. For this work, however, we have a mix of cluster coverages (ranging from $0.35$ R$_{200}$ to $2.11$ R$_{200}$) and if we only consider clusters covered to at least a given radius (e.g. $0.7$ R$_{200}$ to maximize galaxy numbers) and we remove objects without spectroscopy (for plotting on phase-space diagram), we are left with a low number of galaxies (see Figure \ref{fig:tail_hist_rmin} in Appendix~\ref{tad_fdc}, where we inspect the effect of the cluster coverage on the tail-BCG angle distribution). Therefore, we modify the method to use the (projected) radial distribution of the galaxies (instead of phase-space) and the tail direction distribution. Note that because this new method does not use phase-space, we do not have to restrict the sample to the  spectroscopic one. Furthermore, in the model, we now mimic the conditions of our sample, by cropping the extent of the simulated clusters, following the same coverage distribution of the clusters from the observations. Other than these changes, the Bayesian approach follows a similar setup to \citetalias{Smith2022} (section 3.4), where the likelihood is defined in the same manner (but note that here we are using four radial bins instead of four phase-space regions), and we use uniform prior distributions for $r_{1}$, $\delta$, and $t_{2}$, within the same range as defined in table 1 from \citetalias{Smith2022}.

\begin{figure}
    \centering
    \includegraphics[width=0.48\textwidth]{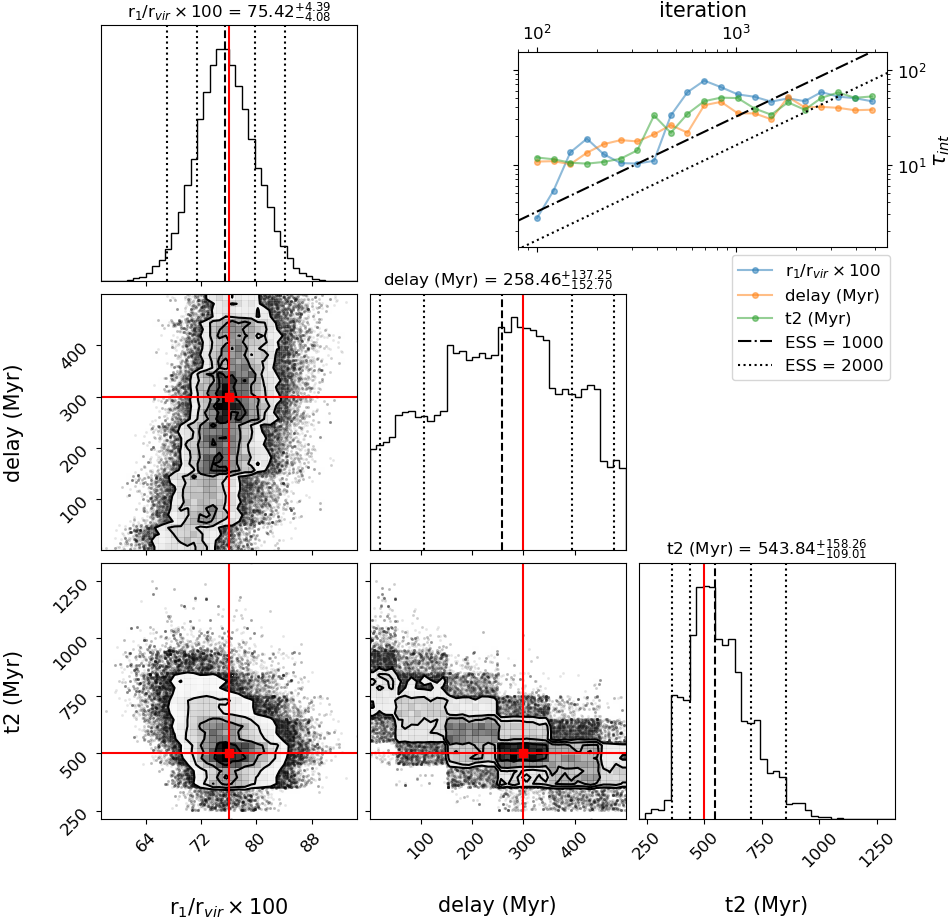}
    \caption{Bayesian parameter estimation results for mock data (as in \citetalias{Smith2022}). Panels are arranged as follows. In the upper right, a convergence monitoring panel is shown for each parameter (See \citeauthor{Shinn2020} \citeyear{Shinn2020} for a more detailed explanation). The panels with grayscale shading and contours are two dimensional PDFs comparing two different model parameters. The upper left, center, and lower right panels are marginalized PDFs of $r_{1}/R_{200}$, $\delta$, and $t_{2}$, respectively. The central vertical dashed lines are the median of the respective distributions, while the surrounding vertical dotted lines show the $68\%$ and $95\%$ credible intervals. The subtitles of the panels provide the median values, and the errors are for the $68\%$ credible interval. The red lines show the input value for each parameter. 
    }
    \label{fig:MCMC_test}
\end{figure}

\begin{figure*}
    \centering
    \includegraphics[width=0.48\textwidth]{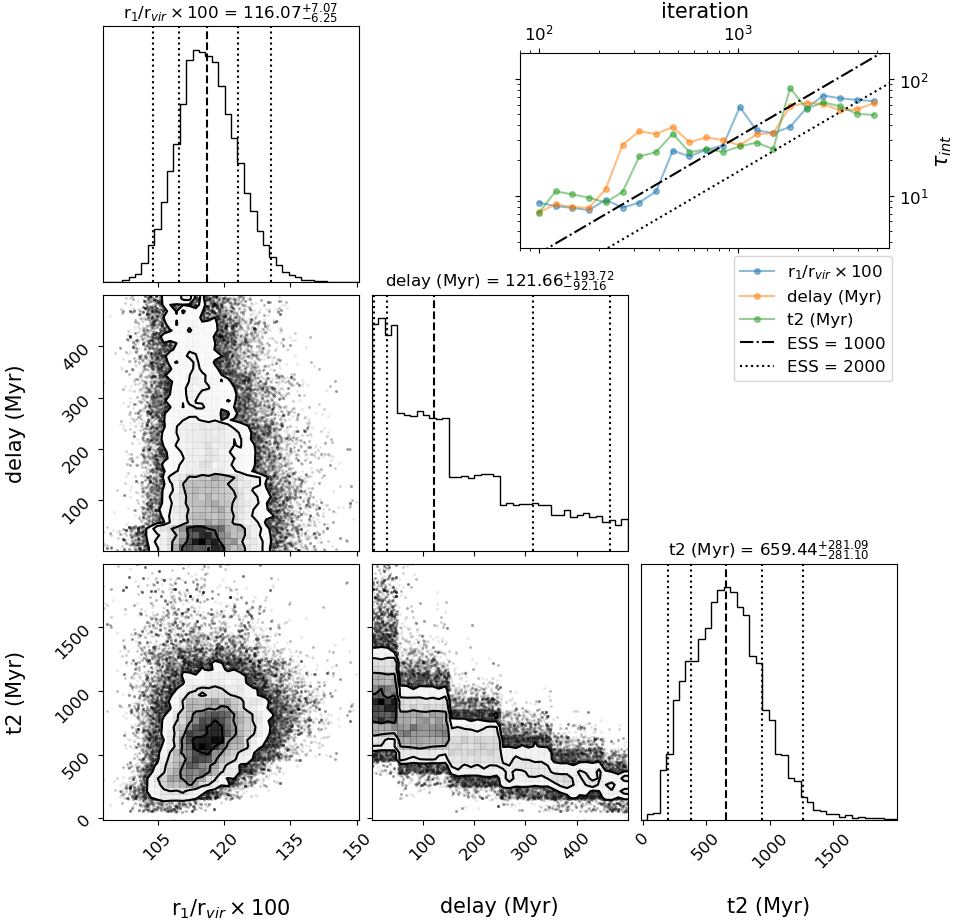}
    \includegraphics[width=0.48\textwidth]{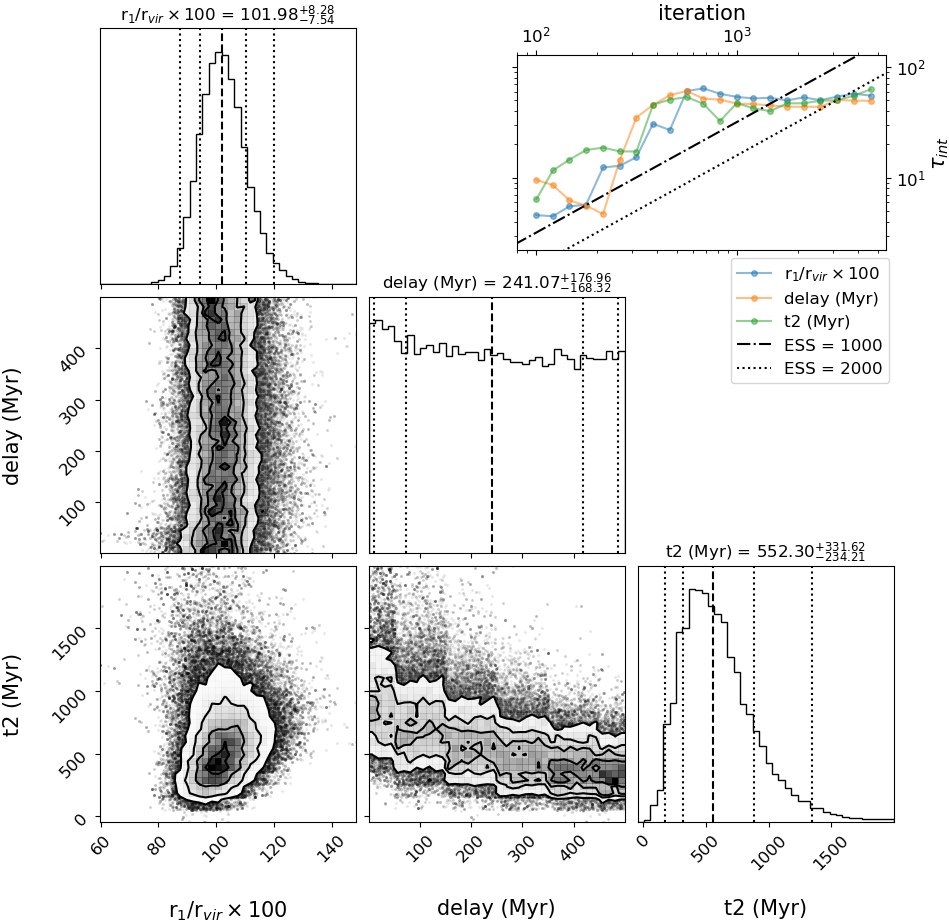}
    \caption{Bayesian parameter estimation results for all tails (left panels) and for only clear confidence tails (right panels). Panels are arranged as in Figure \ref{fig:MCMC_test}. 
    }
    \label{fig:MCMC}
\end{figure*}

To test the new model and modified method in Figure \ref{fig:MCMC_test} we show a mock test of the modified model, using input values of ($r_{1}$, $\delta$, $t_{2}$) $=$ ($76$, $300$, $500$). From here we find that the model can easily reproduce the initial input values within the $68\%$ credible interval, obtaining good constraints on $r_{1}$ and $t_{2}$. The $\delta$ parameter is the only poorly constrained value. This was also seen in \citetalias{Smith2022},  indicating that the results are not strongly sensitive to the value of $\delta$. Nevertheless, the success of the modified method in retrieving the $r_{1}$ and $t_{2}$ parameters of the mock data set provides confidence for running the model on our data and motivates the use of this method on large photometric samples. 

We then ran the model with the observations using galaxies with both marginal and clear tails, and another using only clear tails. The results are presented in Figure \ref{fig:MCMC}. In the case of all tails (left panels), we find median values of $r_{1} = 1.16^{+0.07}_{-0.06}$ R$_{200}$ and $t_{2} = 659^{+281}_{-281}$ Myr, while in the case of only clear tails (right panels), we find $r_{1} = 1.02^{+0.08}_{-0.08}$ R$_{200}$ and $t_{2} = 552^{+332}_{-234}$ Myr. Both results are in fairly good agreement within errors and indicate that the tails are formed very early upon entering the cluster, and disappear shortly after pericenter. It is also expected that clear tails seem to appear a bit further into the cluster, when ram pressure starts to overcome the anchoring force of the galaxies during the first infall into the cluster. They also dissapear a bit earlier. 

\section{Discussion}
\label{sec:discussion}

In this section, we interpret all of our results combined in order to provide a common framework regarding the typical orbits of jellyfish galaxies in clusters, the lifespan of the tails, and the relation between the tail orientation and other properties.

\subsection{Radial orbits favour RPS}

The first key result of our study is the distribution of jellyfish galaxy tail directions relative to the cluster center. Figure \ref{fig:tail_hist} shows a clear preference for tails pointing away from the BCG in our optical jellyfish candidate sample, followed by perpendicular tails, and tails pointing towards. This is consistent with the original \citetalias{Poggianti2016} estimates, which had only a slightly lower fraction of galaxies pointing towards (by $5.5\%$). It is also consistent with other tail directions studies at various wavelengths (e.g. \citeauthor{Chung2007} \citeyear{Chung2007}, \citeauthor{SmithRu2010} \citeyear{SmithRu2010}, \citeauthor{Roberts&Parker2020} \citeyear{Roberts&Parker2020}, \citeauthor{Roberts2021} \citeyear{Roberts2021}). 

If we visualize the orbit of an infalling galaxy radially moving towards the cluster center (Figure~\ref{fig:angle_diagram}), we can expect that, as the galaxy approaches pericenter, it first develops a tail that points away from the BCG, but once it gets near pericenter the tail will change direction into a perpendicular tail (briefly), and after pericentric passage it will point towards the BCG until it dissapears. In this context, the observed distribution of tail-BCG angle (skewed towards high $\theta$ values) can be interpreted as a result of \textit{RPS being more efficient for galaxies following radial orbits on their first infall into the cluster}. In fact many of the previous works have adopted this interpretation. Here we present several additional pieces of evidence supporting this idea: 
\begin{itemize}
    \item Using the models we built to interpret our results, we are able to confirm that most ($\sim 90 \%$) of the galaxies moving towards the cluster center (i.e. with tails pointing away from it) are indeed galaxies on first infall into the cluster that have not reached pericenter yet (see panel B in Figure \ref{fig:simulated_before_after_peri}).
    \item Tails pointing away have the strongest signatures of RPS (see Figure \ref{fig:jclass_hist}). 
    \item RPS candidates with tails pointing away from the cluster center tend to have higher absolute relative velocities with respect to the cluster mean when compared to other tail orientations \citep[see phase-space diagram of Figure~\ref{fig:PPS}, as well as e.g.][]{Jaffe2018}, and higher velocity dispersion outside the cluster core (see VDP in Figure~\ref{fig:VDP}). Given that stripping is proportional to the square of galaxy velocity (see Equation~\ref{eq:G&G}), it is indeed expected that high-speed (radial) first infallers will have the clearest signs of stripping.  Interestingly, near the cluster core we observe a drop in the VDP of the tails pointing away, allowing the perpendicular tails to have a larger velocity dispersion when compared to the away sample. This is not completely unexpected, because galaxies on radial orbits near pericenter can indeed show perpendicular tails. This can happen not only as a projection effect, but because at one point in a radial orbit, a galaxy would be moving perpendicular to the BCG for a short time, and would be in the process of changing tail direction. Therefore, these special cases of perpendicular tails would be an intermediate type between a tail pointing away and towards the cluster center, although our models suggest that overall, the perpendicular population is likely dominated by non-radial orbits (see section \ref{mock_sims}). This can partly explain the observed drop of the away sample, and is why the VDP of all tail orientations near the cluster core are similar within errors since the high velocity cases expected from radial orbits are spread into the three tail orientations when observed near the core, whereas outside the core most infalling galaxies should result in a tail pointing away from the BCG. In addition to this, the shape of the VDPs could be significantly affected by the low number of galaxies in the sample.
    \item The VDP of the jellyfish candidates with no tail (black solid line in Figure~\ref{fig:VDP}) have the lowest velocity dispersion near the cluster center (although not significantly different from the other galaxies within errors). Galaxies entering the cluster with less plunging orbits are indeed less likely to produce significant tails, as they would not be able to reach further into the cluster where the density is higher, nor would they reach high enough velocities for effective RPS. However, at distances larger than $\sim0.5$ R$_{\text{200}}$ they have the second largest velocity dispersion. Therefore, this result could be an indication that these galaxies have a mixture of orbital shapes. We present a detailed analysis of the orbits of RPS candidates by the inversion of the Jeans equation in \citep{Biviano2024}, where we find that orbits of RPS candidates are increasingly radial with distance from the cluster center, from almost isotropic at the center, to very radial at the virial radius. 
\item Finally, the existence of galaxies with tails pointing towards the BCG (with strong RPS features) suggest that at least in some cases the RPS-induced tails in radially infalling galaxies can be visible even after pericentric passage. The constraints on the RPS duration is discussed in Section~\ref{subsec:timescale}. 
\end{itemize}

Not all jellyfish candidates have tails pointing towards or away the cluster center. In fact, \textit{there is a significant fraction of perpendicular tails}. Perpendicular tails can occur in three possible scenarios (ignoring projection effects and inhomogeneities in the ICM): (i) when a galaxy enters the cluster for the first time with a less plunging orbit, (ii) when a galaxy following a radial orbit is near pericenter or apocenter, and (iii) when the initially radial orbits transform into a more circular one. 

Using our models, we show that at the very least half of the galaxies with perpendicular tails entered the cluster on a less plunging orbit, whereas  the other half (non-first infallers) could have transformed their orbits later (see panel B in Figure \ref{fig:simulated_before_after_peri}). When inspecting the distribution of the pericenter distance in our simulated data (Panel C of Figure \ref{fig:simulated_before_after_peri}), we find considerably larger values for perpendicular tails compared to other orientations, confirming that perpendicular tails do not preferentially occur on radial orbits. 

Our observations support the scenario where most of the perpendicular tails correspond to less plunging (circular) orbits. The evidence includes the wide radial distribution of the perpendicular tails sample, with an average at a high clustercentric distance (Figure \ref{fig:radial}), and the VDP of the perpendicular sample, which is high at the inner parts (where radial orbits can more easily cause perpendicular tails), and drops lower than the away sample in the outer parts (Figure \ref{fig:VDP}). 

Finally, galaxies with perpendicular tails have weaker RPS features than galaxies with other tail orientations (Figure \ref{fig:jclass_hist} respectivelly), which confirms the hypothesis that is the radial orbits which causes the strongest stipping effect. 

\subsection{Timescale of the RPS process}
\label{subsec:timescale}

Another important result is the existence of strong stripping features in galaxies with tails pointing towards the cluster (Figures~\ref{fig:tail_hist} and~\ref{fig:jclass_hist}).This finding suggests that tails can survive a galaxy's pericentric passage, but the timescale that a a radially infalling galaxy will show a tail pointing towards the BCG is smaller than that of the one with a tail pointing away (which is the dominant orientation). In other words, the reduced number of galaxies with tails pointing towards relative to tails pointing away suggest that many galaxies will be completely stripped before or shortly after pericentric passage (see Figure~\ref{fig:angle_diagram}).

Our models  indeed suggest that the  abundance of high-confidence tailed galaxies with low $\theta$ could correspond to \textit{a population of galaxies that have recently passed pericenter, but have not completely lost their tails yet (see panel A in Figure \ref{fig:simulated_before_after_peri})).}

The following observations support this scenario: 
\begin{itemize}
    \item The radial distribution of the tailed galaxies, shown in Figure \ref{fig:radial}, shows that tails pointing towards the BCG peak closer to the cluster center when compared with the away sample, further suggesting tails in radially outfalling galaxies do not remain visible for long after pericenter, or have changed direction. 
    \item Figure \ref{fig:VDP} further shows a steep monotonically decreasing VDP from small to large clustercentric distances for galaxies with tails pointing towards the cluster, which is consistent with these objects being outfalling galaxies in radial orbits that are losing speed as they move farther from the center, which will inevitably weaken the RPS process.
\end{itemize}

A caveat worth mentioning is that we expect some contamination from galaxies close to pericentric passage travelling along our line of sight that appear to have tails pointing towards the cluster in projection, but are actually pointing away in 3D \citep[e.g. the extreme case of JO201; ][]{Bellhouse2017}. The impact of projection effects in the tail-BCG angle distribution can be roughly estimated by inspecting the set of simulated galaxies whose tails disappear before pericenter, shown in Figure \ref{fig:simulated_before_after_peri} (panel A). Because the tails are set to disappear before reaching pericenter, we know that the relatively small number of galaxies with low tail-BCG angles seen there can only be caused by projection effects. However the contamination is fairly minor, and the observed distribution in Figure \ref{fig:tail_hist} (and in particular the strong presence of towards tails in clear RPS cases) strongly suggests that not all galaxies are completely stripped (and devoid of visible tails) on first infall, and that RPS can remain effective in producing tails after pericentric passage.

We were able to robustly quantify the lifespan of the tails by combining the tail distribution with the radial distribution of the jellyfish candidates, and  using the Bayesian parameter estimation method showcased in Figure \ref{fig:MCMC}.  finding that tails are able to survive after pericenter for $\sim 38 \%$ of the total lifetime of the tails. We also note that this is often not enough time to allow for second passages while retaining (or recovering) a visible tail (see panel B in Figure \ref{fig:simulated_before_after_peri}), although it should be possible in a relatively small number of cases.

The average overall lifetime of the tails based on the constraints obtained in Figure \ref{fig:MCMC} yields values of $\Delta t = 1.73\pm0.48$ Gyr when using all tails, and $\Delta t = 1.44\pm0.44$ Gyr when using only clear tails. These are relatively short times when compared with the total RPS timescale predicted for the gas in the IllustrisTNG simulations from \cite{Rohr2023}, where they find a range between $1.5 - 8$ Gyr. However, they mention the peak of the gas stripping occurs within $0.2-2$ R$_{200}$ and lasts less than $2$ Gyr. Since our sample is typically found within less than $2$ R$_{200}$ our estimates appear to be in good agreement with their results, in which case the average period of visibility of optical tails would coincide with the moment in which the stripping is most effective. Although, our results might be regarded as a lower limit since we are limited by the maximum coverage of the sample, and it may be that tails appear at even larger distances. Expanding the search of optical tails at greater distances would allow to explore the possibility of star formation activity prior to the peak of gas stripping.

From an observational perspective on the gas stripping, detailed HI studies of galaxies experiencing RPS in nearby clusters are also compatible with our constraints: In the Fornax cluster \cite{Loni2021} concludes that the total neutral atomic gas content should be lost within a crossing time of $\sim 2$ Gyr, and in the Hydra cluster \cite{Wang2021} finds weak RPS examples starting as far as $\sim 1.25$ R$_{200}$ in projection, consistent with our value of $r_{1} \sim 1.16$ R$_{200}$. Furthermore, galaxies with surviving HI tails are seen past pericenter in the Hydra cluster \citep{Hess2022}, supporting a positive value for $t_{2}$.

We note that our results yield notably larger values for $r_{1}$ and longer values of $t_{2}$ when compared with the radio continuum results from \citetalias{Smith2022}, where $r_{1} \sim 0.76$ R$_{200}$ and $t_{2} \sim 480$ Myr. This would indicate that optical features in the tail appear sooner and last longer than radio continuum features such as the synchrotron emission caused by supernova explosions. However, in order to carry-out a proper comparison between the two wavelengths we would need to apply the same method for both samples since in this work we are using the radial profiles, rather than the position in phase-space. Additionally, we would need to account for the differences in the samples, such as selection biases and physical properties of the clusters and galaxies. We note that the smaller cluster coverage of the radio continuum sample ($\sim1.05$ R$_{200}$) would not be responsible for the difference since galaxies that form a tail at a larger distance would still have an observable effect once they enter further into the cluster, and galaxies at larger distances can appear inside the coverage when projected, which can then be captured by the model. A future study directly comparing both results could add valuable insight on the physical processes and the effects RPS has on different galaxy components.

\subsection{Dependance on galaxy and cluster properties}

We expect that the impact of RPS and hence the timescale of tail appearance/dissapearance will depend on cluster and galaxy mass \citep[see][]{Jaffe2018,Gullieuszik2020}, but we do not find significant differences in host mass, galaxy mass or color for most of the different tail orientations. The only exception is the statistically different and bluer color distribution of the perpendicular tails, which could be an indication of less effective quenching, further reinforcing the possibility of less effective RPS in non-radial orbits.  \citetalias{Smith2022} found mildly larger differences between galaxies with different orientations\footnote{We note that the towards and away samples in \citetalias{Smith2022} are defined as the tail-BCG angles $\theta < 90$ degrees and $\theta > 90$ degrees, respectively (i.e. they do not use a perpendicular subsample)}, but these differences were still not too significant according to a KS test (p-values range from 0.13 to 0.34). With upcoming large-area surveys of JF galaxies in progress, we will be able to test this further.

Where we do find a difference is when separating clusters by dynamical state. 
The connection between tails and shape of the orbits made in this work is only reasonable when we consider relaxed environments, with a mostly isotropically increasing density from the inner to the outer parts of the cluster. For this reason most of our analysis excluded interacting clusters. However, in Figure \ref{fig:tail_hist_int_cl} we look at the interacting clusters separatelly, and find more randomly distributed tail orientations than in Figure \ref{fig:tail_hist}, although still showing some preference for tails with high tail-BCG angles. Possible factors that could enhance or suppress RPS signatures are passing shock fronts \citep{Rawle2014}, which could alter the velocities of galaxies relative to the medium and/or the density of the medium by moving the material farther or closer to the galaxies. Unfortunately, with the limited sample we have for candidates in interacting clusters, we can only provide speculative interpretations. Better statistics and a more detailed analysis would be needed to have a good understanding of the effect of interacting clusters on galaxy tails since their properties could widely vary for different interacting clusters (see e.g. \citeauthor{Lourenco2023} \citeyear{Lourenco2023}, \citeauthor{Piraino2024} \citeyear{Piraino2024}).

\subsection{Caveats}
Although our study uses the most extensive sample of optical jellyfish candidates in the literature, it has some limitations. One is that a small fraction of clusters in the sample covers more than $2$ R$_{200}$. Hence, we could be missing a fraction of galaxies that produce tails very far from the cluster, since we do find some examples at these distances in the few clusters that reach that far (see also \citeauthor{Piraino2024} \citeyear{Piraino2024}). The inhomogeneous coverage of clusters  makes the interpretation of results more challenging, as reflected by the different tail angle distributions seen for different apertures. This adds a bias against recently infalling galaxies for every cluster with a small coverage. We note however that we overcome this limitation in our analysis as the Bayesian analysis model used in this paper takes into account the different cluster coverages. Any degeneracy in the model parameters is captured in the posterior distributions for $r_1$ and $t_2$.

Another limitation of our study is the sample size (despite having the current largest sample for this work), especially after cleaning and sub-dividing the sample in different tail orientations, lowering the statistical significance of some of our results. The VDP (which is further limited by spectroscopic members) is the most affected in this regard, showing great uncertainty in the results obtained. However, with the continuously growing samples of jellyfish galaxies, we expect to find opportunities to repeat this study in the near future to further refine our results.
\\

Overall our combined results indicate that RPS is an efficient and fast-acting process affecting galaxies as they cross the intracluster medium for the first time, preferentially or more significantly on radial orbits.

\section{Summary and conclusions}

In this work we measured and studied the projected tails of optically selected ram pressure stripping (jellyfish) candidates. Tails can appear when gas is stripped from the main body of a galaxy while it plunges into the dense Intracluster Medium (ICM). Within these tails, stars have the opportunity to form, emitting light that excites stripped gas. Consequently, stripped tails become detectable in optical wavelengths for a certain period of time. Because tails in jellyfish galaxies are expected to point opposite to their direction of motion,  we can use them to reconstruct the orbital histories of galaxies in clusters and the timescale of the stripping features.

We use the largest optically-selected jellyfish candidate sample in local clusters known to date, taken from the works of \citetalias{Poggianti2016} and \citetalias{Vulcani2022} using observations from the WINGS and OmegaWINGS surveys. This sample comprises $379$ jellyfish candidates in total, but reduces to $227$ when removing galaxies with signs of gravitational interactions and/or with a redshift locating them outside the targetted clusters. 
Up to seven classifiers visually inspected the broad-band optical images of the jellyfish candidates to determine the tail directions and confidence of the tails, from which we then took an average value based on the directions that agreed within a margin of $45$ degrees. A good agreement was found between the classifiers, who were able to define a tail angle (relative to the cluster center) for $71\%$ of the jellyfish candidate sample studied. To test the accuracy of our results we  compared the tail directions measured in broad-band optical images with those from H$\alpha$ emission for a subsample of galaxies with MUSE data, finding a good agreement (within $45$ degrees) in $\sim70\%$ of the cases. This comparison provides support to the use of broad-band optical images in the studies of ram pressure stripping features when narrow-band or integral field spectroscopy are not available.

 We obtained the following results from the analysis of the observed tail directions in the cluster jellyfish candidates: 
\begin{itemize}
    \item The distribution of jellyfish tail directions with respect to the cluster center (defined by the BCG) shows a preference for tails pointing away from  the cluster. The distribution however, spans all the range of tail-BCG angles, monotonically decreasing from larger to smaller angles, such that $33\%$ of the galaxy tails point away ($\theta \ge$ 135 degrees),  $49\%$ point perpendicular ($45 \:\text{degrees} \leq \theta < 135 \:\text{degrees}$), and $18\%$ point towards the BCG ($\theta < 45$ degrees).
    \item  The strongest stripping signatures, as defined by JClass in \citetalias{Poggianti2016},  are present in galaxies with tails pointing away  or towards the cluster center. These have $38\%$ and $\sim 40\%$ of cases with JClass $>2$ respectivelly. In contrast, perpendicular tails only have $20\%$ of galaxies with JClass $>2$. The most convincing cases of stripping (JClass$=5$) are almost exclusivelly galaxies with tails pointing away from the cluster.

    \item In projected position vs. velocity phase-space diagram,  galaxies with tails pointing away display the highest overall velocities (and velocity dispersion profile, VDP). Near the cluster core the towards and perpendicular tail orientations also show a high velocity dispersion. The VDP of the towards sample decreases monotonically to larger clustercentric distance. Most notably, the galaxies with perpendicular tails have the lowest overall VDP. We also find that galaxies with clear tails prefer high velocities near the cluster core. 
    
    \item The radial distribution of the larger sample (with or without spectroscopy) shows a distribution that peaks at $\sim0.5$ R$_{200}$ for tails pointing away or perpendicular to the BCG, while tails pointing towards show a peak closer to the center. We find a typical average distance that tends to be around $0.64$ R$_{200}$ (depending on the tail orientation and clarity of the tails). 

    \item The properties of the galaxies (mass and color) or host clusters (host mass and mas ratio) are not significantly different for the galaxies that have tails pointing towards or away. Perpendicular tails were the most different with respect to the other tail orientations, showing slight preferences to be bluer than galaxies with tails pointing away from the cluster center. Finally, when splitting clusters by dynamical stage, we found that interacting clusters display a flatter distribution of the tail-BCG angle relative to regular clusters.
\end{itemize}

Our results are consistent with the formalism by \cite{Gun&Gott1972}, which predicts stronger ram-pressure in galaxies falling in dense ICM at high speeds (see equation~\ref{eq:G&G}). It also confirms previous claims based on observations and/or simulations \citep[e.g.][]{Jaffe2015,Jaffe2018,Smith2022,Biviano2024} that infalling galaxies on radial orbits (in fairly regular clusters) likely experience stronger ram pressure stripping.

To test this hypothesis and deepen our understanding in the ram-pressure stripping process and its consequences in cluster galaxies, we compared the observational results with simple modeling folded into N-body cosmological dark matter only simulations. 

The simulated data indicates that the observed preference for tails pointing away from the cluster (together with their position and velocity profiles and JClass) is expected in galaxies experiencing RPS as they fall into the cluster for the first time on fairly radial orbits. The less common (but still present) population of galaxies with tails pointing towards the cluster on the other hand, correspond to these radially infalling galaxies that have already passed pericenter and have not yet lost their tail completely.  Finally, galaxies with perpendicular tails (and weaker stripping features) are consistent with less radial orbits.

We further apply the Bayesian analysis method introduced by \citetalias{Smith2022}  to obtain quantitative constraints on the lifespan of optical tails. We adapted the model to suit the coverage distribution of our sample and modified the method to use the radial distribution of jellyfish candidates instead of the phase-space coordinates to compensate for the small spectroscopic sample. We find that optical tails appear for the first time at a clustercentric distance of $r_{1} = 1.16^{+0.07}_{-0.06}$ R$_{200}$ and disappear $t_{2} = 659^{+281}_{-281}$ Myr after pericenter, confirming ram pressure stripping is an important and imminent physical mechanism transforming galaxies soon after they enter a galaxy cluster for the first time. And that (optical) jellyfish tails can remain visible  after pericentric passage.

In summary, our combined results are consistent with ram pressure stripping being more effective for galaxies falling into the cluster preferentially on radial orbits, and suggest the optical tails in jellyfish galaxies (lit up by star formation happening in the stripped gas) are somewhat short lived, but can be visible even after the pericentric passage. 

Using the novel method introduced by \citetalias{Smith2022} we also find that ram pressure stripping features (tails) typically start to appear just beyond R$_{200}$ and can be visible for a considerable amount of time after pericentric passage. 

Follow up work of this study to deepen our understanding of the stripping process can involve:  analyzing an even larger and homogeneous sample from e.g. citizen science efforts (see Zooniverse project "Fishing for Jellyfish galaxies"\footnote{https://www.zooniverse.org/projects/cbellhouse/fishing-for-jellyfish-galaxies}); doing a more detailed analysis of the results obtained in this paper, such as splitting by cluster and galaxy properties; and comparing with stripped galaxies at other wavelengths. This is going to broaden our understanding of ram pressure stripping and the effect this mechanism has on the star formation of cluster galaxies.

\section*{Acknowledgements}
Y.J. acknowledges financial support from ANID BASAL project No. FB210003 and FONDECYT Project No. 1241426. RS and YJ acknowledge support from FONDECYT Regular No. 1230441. AB acknowledges financial support from the INAF mini-grant 1.05.12.04.01 {\it "The dynamics of clusters of galaxies from the projected phase-space distribution of cluster galaxies"}. A.C.C.L thanks for the financial support of the National Agency for Reaserch and Development (ANID) / Scholarship Program / DOCTORADO BECAS CHILE/2019-21190049. K.K. acknowledges full financial support from ANID through FONDECYT Postdoctrorado Project 3200139, Chile. This project has received funding from the European Research Council (ERC) under the Horizon 2020 research and innovation programme (grant agreement N. 833824). JPC acknowledges financial support from ANID through FONDECYT Postdoctorado Project 3210709. GGT is currently supported by the German Deutsche Forschungsgemeinschaft (DFG) under Project-ID 445674056 (Emmy Noether Research Group SA4064/1-1, PI Sander) and Project-ID 496854903 (SA4064/2-1, PI Sander). F.P.C. acknowledges financial support from Dirección de Postgrado (Universidad Técnica Federico Santa María, Chile) through Becas Internas para Doctorado y Magíster Científico-Tecnológicos.
\section*{Data Availability}

The tail direction measurements of the galaxies presented in this paper are provided in the online version of this paper.






\newpage
\appendix 

\section{Tail angle measurements}

\subsection{Comparisson between classifications}
\label{ap:comparisson}

\subsubsection{Comparison between classifiers}

To 
quantify the agreement in all the classifications, we computed the scatter of the tail angles obtained for each galaxy that had at least 2 classifiers with a confidence level greater than $0$ (i.e. with a visible tail). For this, we used the definition of the standard deviation given by

\begin{equation}
    \centering
    \sigma_{\text{angle}} = \sqrt{\frac{\sum{\left(\Delta \theta_{i}\right)^{2}}}{N}}\text{,}
    \label{eq:Cstd}
\end{equation}

where $N$ is the number of classifiers, and $\Delta \theta_{i}$ is given by

\begin{equation}
\Delta \theta_{i} = \begin{cases}
\mid \theta_{i} - \overline{\theta} \mid,& \text{if } \mid \theta_{i} -\overline{\theta}\mid \leq 180^{\text{o}} \\
360^\text{o} - \mid\theta_{i} - \overline{\theta}\mid,  & \text{otherwise}
\end{cases}
\end{equation}

where $\theta_{i}$ is the tail angle obtained by the $i$th classifier, and $\overline{\theta}$ is the circular average of the angles given by the $N$ classifiers. Note that we are using the circular average and this definition of $\Delta \theta_{i}$ to account for the fact that angles are cyclic quantities. The distribution of standard deviations can be seen Figure~\ref{fig:std_hist}, where we find an average standard deviation of $\sim32$ degrees.

\begin{figure*}
    \centering
    \includegraphics[width=0.99\textwidth]{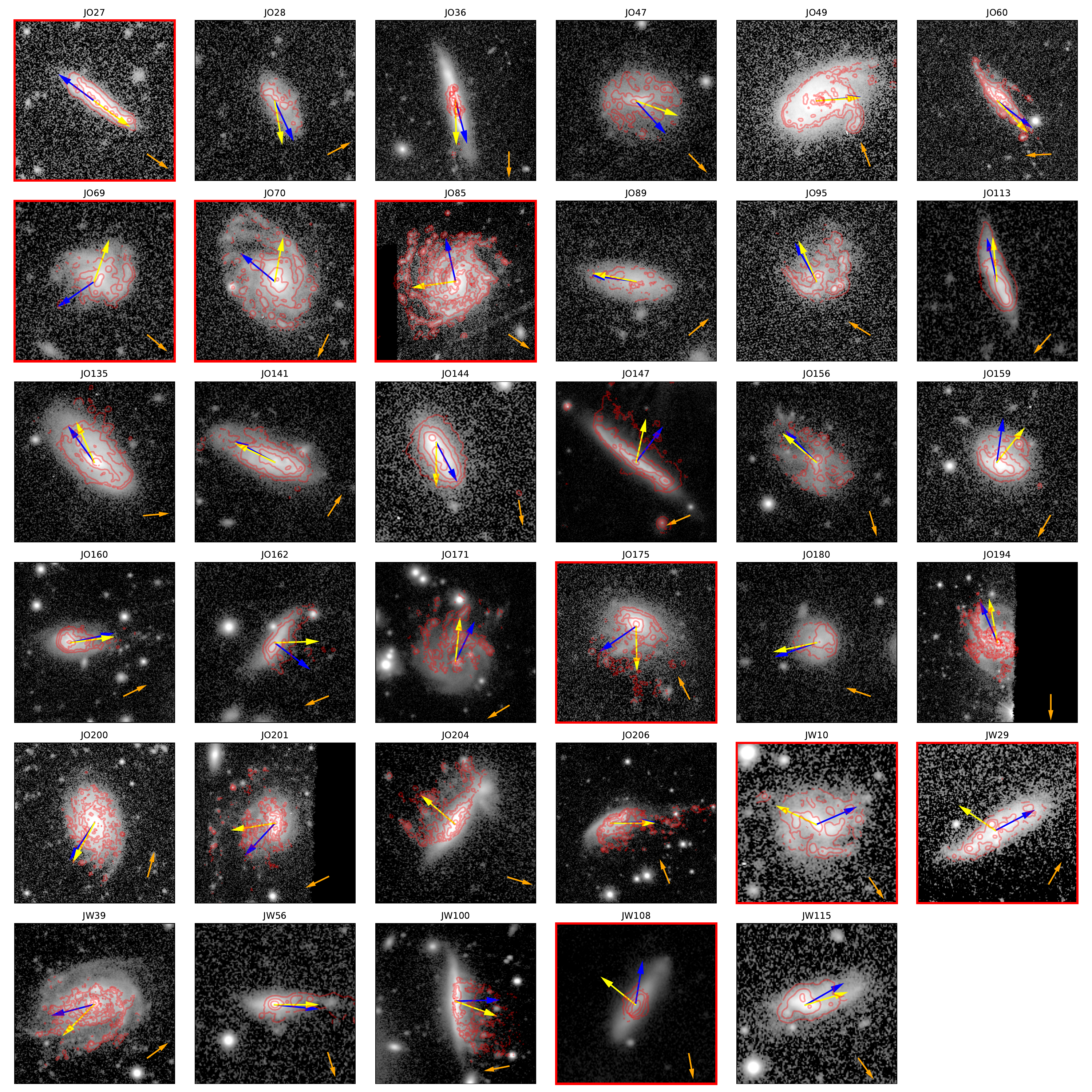}
    \caption{B-band images from WINGS/OmegaWINGS of the $35$ galaxies with tails in both optical and H$\alpha$ emission (red contours), with arrows representing the tail directions in the optical (blue arrows) and in H$\alpha$ (yellow arrows). The orange arrow at the bottom right corner points in the direction of the BCG. Images with red axes highlight cases where tail measurements have a difference greater than 45 degrees.}
    \label{fig:P16vsGASPplots}
\end{figure*}

\subsubsection{Comparison between broad-band optical vs.  H$\alpha$ tails}

To test how reliable are our tail measurements based in broad-band optical images, we compared them with tail measurements done using H$\alpha$ maps from MUSE data, for a subsample of 47 galaxies within our sample that were observed by the GASP survey. Since H$\alpha$ emission usually provides a clearer tracer of the stripped gas tails than broad-band optical images, comparing our results with tail directions using H$\alpha$ provides a good way of further testing the accuracy of the results. The H$\alpha$-based tail directions were visually measured based on the asymmetry of the H$\alpha$ emission with respect to stellar contours.

The classification of galaxies in the GASP sample yielded $41$ galaxies with confirmed tails in H$\alpha$, out of which $35$ also have visible tails in the optical. Note that this difference could be due to faint H$\alpha$ tails not being easily detected in broad-band optical images. However, when tails are seen in both cases we would expect to obtain similar tail directions, such that any deviation can then be interpreted as a systematic error arising from our methodology, which could be caused by the increased difficulty of accurately classifying broad-band optical tails. Figure \ref{fig:P16vsGASPplots} presents optical images of the galaxies for which the optical (blue) and H$\alpha$ (yellow) tail directions were compared. There is a good agreement overall. In total, $30\%$ of the galaxies have a discrepancy greater than 45 degrees, which translates into an agreement of $70\%$. Furthermore, the average difference is $34.9$ degrees, with a standard deviation of $44.3$ degrees\footnote{The  mean and standard deviation of the angle differences here are computed in the standard way, with absolute value ranging from $0$ to $180$ degrees.} (both below our threshold of $45$ degrees). 

When inspecting the $8$ galaxies with large tail angle discrepancies, we find that $4$ of them are unwinding galaxies. This is unsurprising as these are typically spiral galaxies seen face on and stripped along the line of sight \citep[see e.g.][]{Bellhouse2017}, making the (projected) stripping direction harder to define. The other $4$  galaxies that are not unwinding have different reasons for the discrepancy: In the case of JW10, JW29, and JW108, the difference is caused due to the clear amount of the observable debris in the H$\alpha$ maps, which are not as easily observed (or not observed at all) in the optical broad-band images. However, in the case of JO27, which has the largest discrepancy with each arrow pointing in the opposite direction of one another, we do not particularly find a clear indicator of the tail direction from the H$\alpha$ emission, nor from the optical image (classified with marginal confidence). An argument could be made for either of the two directions or even for the non-existence of a tail. Therefore, this is a rare example where a discrepancy with H$\alpha$ does not necessarily means the optical tail direction is wrong.

In conclusion, when comparing broad-band optical versus H$\alpha$ measurement of tail direction in jellyfish candidates, we find that, although tails are more clearly visible through their ionized gas, there is a good agreement.

\subsection{Tail-BCG angle distribution in merging clusters}
\label{ap:merging-cl}

\begin{figure}
    \centering
    \includegraphics[width=0.5\textwidth]{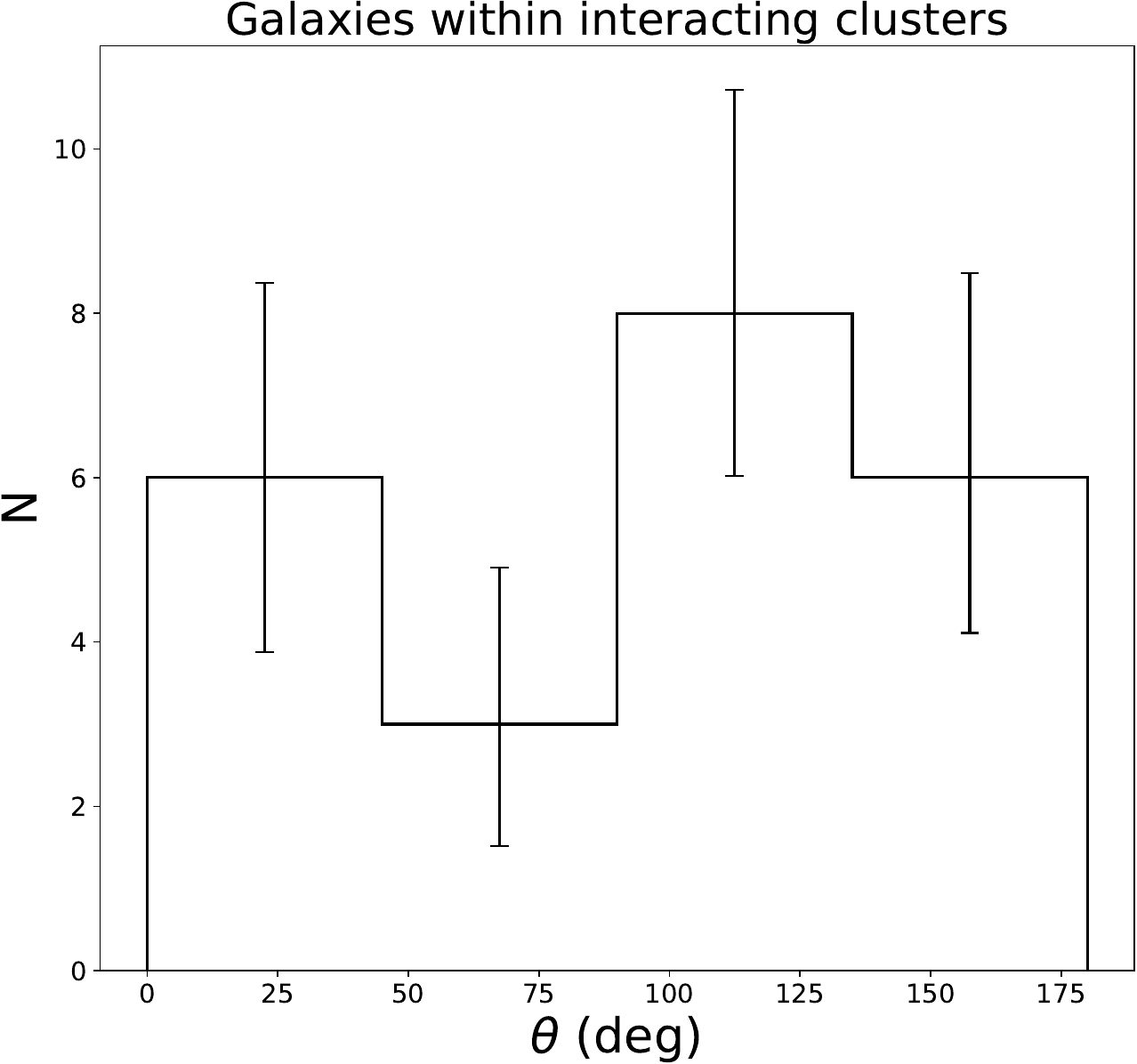}
    \caption{Jellyfish tail angle distribution for galaxies within interacting clusters. This sample is excluding confirmed non-members and interacting galaxies. We do not further divide the sample into more confident classifications (as in Figure \ref{fig:tail_hist}) since the sample gets significantly reduced. Error bars were computed using bootstrapping.
    \label{fig:tail_hist_int_cl}}
\end{figure}

In  Figure~\ref{fig:tail_hist} (and the rest of the paper) we excluded galaxies in merging clusters (along with interacting and non-member galaxies) to have the cleanest possible tail-BCG angle distribution. Here we take a dedicated look at the tail directions in jellyfish candidates inside merging clusters.

Figure \ref{fig:tail_hist_int_cl} shows that the distribution of tail-BCG angles within interacting clusters is  much flatter than the one found in regular systems (Figure~\ref{fig:tail_hist}). This supports the notion that galaxies within unrelaxed clusters might be subject to particular conditions that alter the orbits of the galaxies and/or the medium surrounding them, which could have an effect on both the effectiveness of RPS and in the direction of the tails. 

\subsection{Tail-BCG angle distribution for different coverages}
\label{tad_fdc}

\begin{figure}
    \centering
    \includegraphics[width=0.48\textwidth]{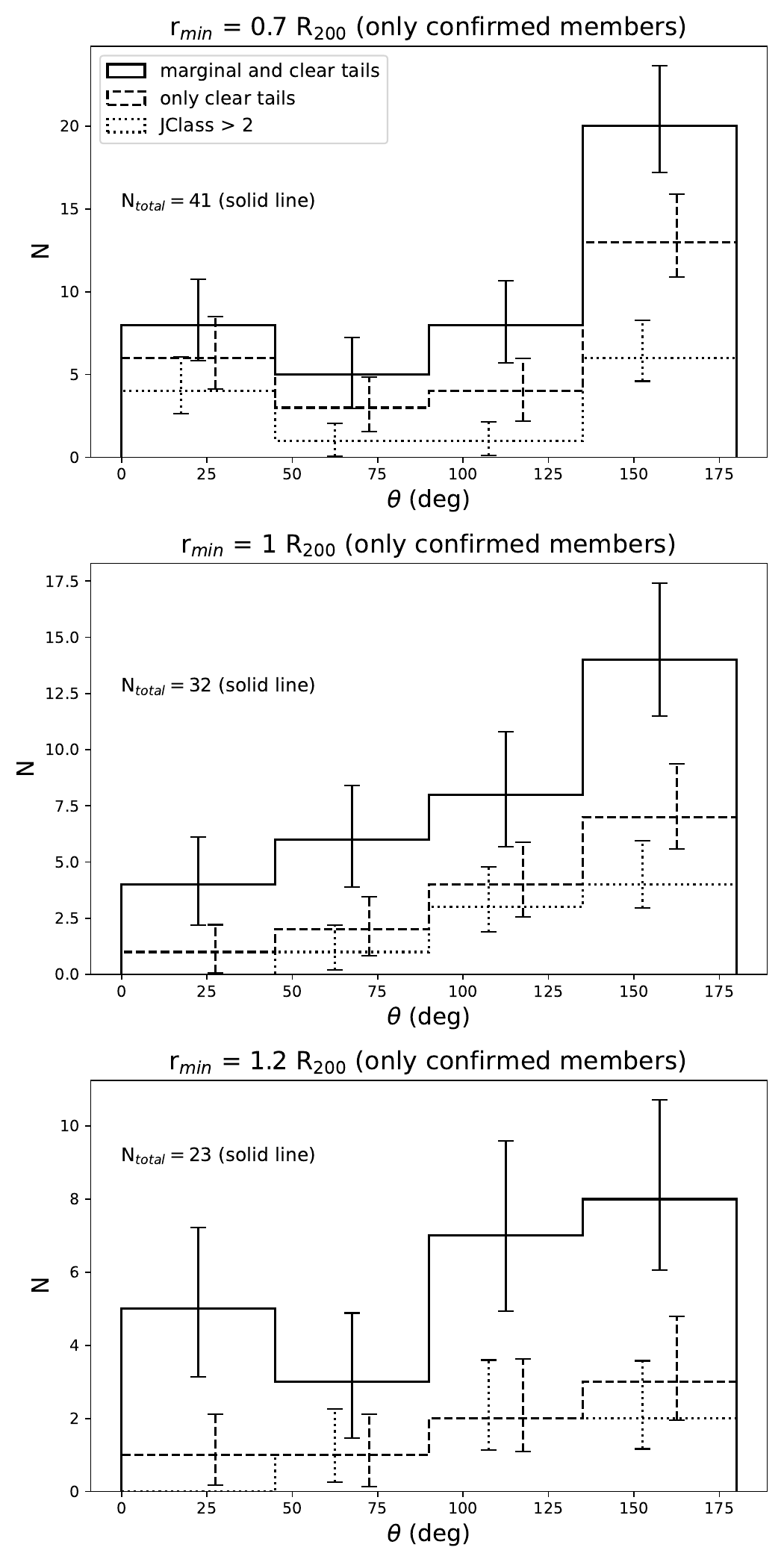}
    \caption{Jellyfish tail-BCG angles for different $r_{\text{min}}$ (top: $0.7 R_{200}$; middle: $1 R_{200}$; bottom: $1.2 R_{200}$). We only considered confirmed cluster members. The sub-samples of different confidence are defined as in Figure \ref{fig:tail_hist}. Error bars were computed using bootstrapping. All plots exclude gravitationally interacting galaxies and galaxies in interacting clusters.}
    \label{fig:tail_hist_rmin}
\end{figure}

We have presented the tail distributions using all jellyfish candidates from WINGS and OmegaWINGS combined, including candidates from both \citetalias{Poggianti2016} and \citetalias{Vulcani2022}. However, not all clusters have the same observational coverage as they vary in mass and redshift. Furthermore, not all WINGS clusters were observed by OmegaWINGS, which had a significantly wider field of view. So in order to fairly combine the tail measurement results of different clusters we limit the sample to clusters that have observations covering up to a given minimum radius $r_{\text{min}}$ and only consider galaxies within a circular aperture of this radius. Ideally, we would like $r_{\text{min}}$ to be larger than $R_{200}$ but when imposing this constraint the sample decreases significantly. We therefore consider different values of $r_{\text{min}}$ to test if the tail angle distribution changes significantly. 

In Figure~\ref{fig:tail_hist_rmin} we show the galaxy tail-BCG angle distribution for the spectroscopically confirmed cluster members in our sample of jellyfish candidates considering the 3 different values of $r_{\text{min}}$ defined above.

When looking only at the cores of clusters ($r<0.7 R_{200}$; upper panel in Figure~\ref{fig:tail_hist_rmin}) we obtain a similar double-peaked distribution to the one in Figure \ref{fig:tail_hist} for JClass $>2$ and clear tails (dotted and dashed histograms), with a clear peak at high tail angles (i.e. tails pointing away from the cluster), and a small secondary peak at low angles (not significant enough in most cases), for tails pointing towards. If we increase $r_{\rm min}$ to $1 R_{200}$ (middle panel in Figure~\ref{fig:tail_hist_rmin}) or even $1.2 R_{200}$ (bottom panel) the distribution does not show a second peak at low angles, at least for clearly tailed galaxies. Lastly, for the largest $r_{\rm min}$    perpendicular tails are more common. Note however that statistics become poorer at increasing $r_{\rm min}$, and that according to a KS test, none of the apparent differences between the distribution in the upper panel and those in the middle and lower panels are significant (p-values of $0.99$ and $0.71$, respectively). We tried to improve statistics by considering all galaxies irrespective of whether they had a spectra or not (not shown), but the results did not change significantly. Larger samples of jellyfish candidates covering a wide area around clusters are needed. 

\bsp	
\label{lastpage}
\end{document}